\documentclass[pre,aps,twocolumn]{revtex4}
\usepackage{epsfig}
\usepackage{bm}
\newcommand{\B}[1]{{\bm{#1}}}

\usepackage[latin1]{inputenc}
\newcommand{\beq}{\begin{equation}}
\newcommand{\eeq}{\end{equation}}
\newcommand{\bea}{\begin{eqnarray}}
\newcommand{\eea}{\end{eqnarray}}
\begin{document}
\title{Void Formation and Roughening in Slow Fracture}
\author{Itai Afek, Eran Bouchbinder, Eytan Katzav, Joachim Mathiesen
and Itamar Procaccia}
\affiliation{Dept. of Chemical Physics, The Weizmann Institute
of Science, Rehovot 76100, Israel}
\begin{abstract}

Slow crack propagation in ductile, and in certain brittle materials,
appears to take place via the nucleation of voids ahead of the crack
tip due to plastic yields, followed by the coalescence of these
voids. Post mortem analysis of the resulting fracture surfaces of
ductile and brittle materials on the $\mu$m-mm and the nm scales
respectively, reveals self-affine cracks with anomalous scaling
exponent $\zeta\approx 0.8$ in 3-dimensions and $\zeta\approx 0.65$
in 2-dimensions. In this paper we present an analytic theory based
on the method of iterated conformal maps aimed at modelling the void
formation and the fracture growth, culminating in estimates of the
roughening exponents in 2-dimensions. In the simplest realization of
the model we allow one void ahead of the crack, and address the
robustness of the roughening exponent. Next we develop the theory
further, to include two voids ahead of the crack. This development
necessitates generalizing the method of iterated conformal maps to
include doubly connected regions (maps from the annulus rather than
the unit circle). While mathematically and numerically feasible, we
find that the employment of the stress field as computed from
elasticity theory becomes questionable when more than one void is
explicitly inserted into the material. Thus further progress in this
line of research calls for improved treatment of the plastic
dynamics.

\end{abstract}
\maketitle

\section{Introduction}
\begin{figure}
\centering \epsfig{width=.45\textwidth,file=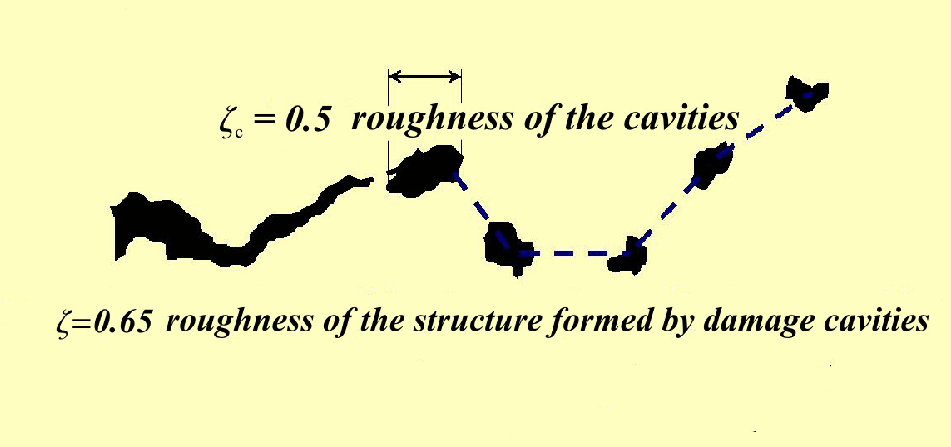}
\caption{The fracture scenario suggested in \cite{99BP}. This
scenario had been documented in detail in corrosive glass
fracture, and also more recently in the fracture of paper
\cite{03Cil}.} \label{Yield}
\end{figure}

In this paper we expand on the results of a recent Letter
\cite{04BMP} in which a model was proposed for slow crack
propagation via void formation ahead of the crack due to plastic
yields. Here `slow' means propagation velocity considerably
smaller than the Rayleigh wave speed. The model was motivated by some
quantitative studies of fracture surfaces, which reveal
self-affine rough cracks with two scaling regimes: at small length
scales (smaller than a typical cross-over length $\xi_c$) the
roughness exponent is $\zeta\approx 0.5$, whereas at scales larger
than $\xi_c$ the roughness exponent is $\zeta\approx 0.8$. The
second scaling regime is seen to have an upper cut-off $\xi$ known
as the correlation length. Such measurements were reported first
for ductile materials (like metals) where $\xi_c$ is of the order
of 1 $\mu$m \cite{84MPP, 97B}, and more recently for brittle
materials like glass, but with a much smaller value of $\xi_c$, of
about 1 nm \cite{03C}. Similar experiments conducted on
2-dimensional samples reported rough cracks with large-scale
exponents $\zeta\approx 0.65\pm0.04$ \cite{93KHW,94EMHR,03SAN}.
The exponent $\zeta\approx 0.5$ is characteristic of uncorrelated
random surfaces, but higher exponents indicate the existence of
positive correlations \cite{Feder}; naturally, the experimental
discovery of such correlated ``anomalous" exponents attracted
considerable interest with repeated attempts to derive them
theoretically \cite{91HHR,03HS,97REF}. Ref. \cite{04BMP} presented
a quantitative model for self-affine fracture surfaces based on
elasticity theory supplemented with considerations of plastic
deformations. Focussing on infinite 2-dimensional materials, Ref.
\cite{04BMP} followed the qualitative picture presented recently
in \cite{99BP}, see Fig. \ref{Yield}. In this picture there exists
a ``process zone" in front of the crack tip in which plastic yield
is accompanied by the evolution of damage cavities. A crucial
aspect of this picture is the existence of a typical scale,
$\xi_c$, which is roughly the distance between the crack tip and
the first void, at the time of the nucleation of the latter. The
voids are nucleated under the influence of the stress field
$\sigma_{ij}(\B r)$ adjacent to the tip, but not {\em at} the tip,
due to the existence of the plastic zone that cuts off the purely
linear-elastic (unphysical) crack-tip singularities. The crack
grows by coalescing the voids with the tip, creating a new stress
field which induces the nucleation of new voids. In the picture of
\cite{99BP} the scale $\xi_c$ is also identified with the typical
size of the voids at coalescence. A consequence of this picture is
that the roughening exponent $\zeta\approx 0.5$ corresponds to the
surface structure of individual voids, whereas the large-scale
anomalous exponent has to do with the correlation between the
positions of different voids that coalesce to constitute the
evolving crack. Ref. \cite{04BMP} provided first a theory for the
scale $\xi_c$ and second, demonstrated that the positions of
consecutive voids are positively correlated. In this paper we
amplify on these results.

In Sect. \ref{onevoid} we discuss again the model of Ref.
\cite{04BMP}, expand the theoretical presentation, and test the
robustness of the roughening exponent that this model predicts.
Since the roughening exponent found is larger than 0.5, this
indicates the existence of positive correlations between crack
increments. To illuminate these long range correlation we
demonstrate in Sect. \ref{long range} that the analytic structure
of the theory dictates the existence of power-law correlations
between height fluctuations in the crack and the value of the mode
II stress intensity factor $K_{_{\rm II}}$ at the tip. In Sect.
\ref{twovoids} we construct a model with two voids ahead of the
crack. After setting up the problem, we describe in some detail
the mathematical apparatus that employs conformal maps from the
annulus to the doubly connected region of a crack with a void
which allows the computation of the stress field around such a
configuration. In Sect. \ref{twovoidsres} we present results of
the two-void model, and discuss the relevance of the results to
the growth and roughening of cracks. The conclusion is that since
the details of plastic deformations are not well understood the
physics of crack growth is better described by the one-void model
than the two-void model; the stress field computed for a crack
with one void ahead is physically acceptable as long as elasticity
theory is relevant, but when the first void appears due to plastic
events a correct determination of the stress field should include
a better handling of the plastic zone. This must await future
improvement of our understanding of plastic dynamics.

\section{One-void model}
\label{onevoid}

\subsection{The plastic zone and void nucleation}
\label{vonmises} A simple model for $\xi_c$ can be developed by
assuming the process zone to be properly described by the Huber-von
Mises plasticity theory \cite{90Lub}. This theory focuses on the
deviatoric stress $s_{ij}\equiv \sigma_{ij} -\case{1}{3}{\rm Tr}
\B\sigma\delta_{ij}$ and on its invariants. The second invariant,
$J_2\equiv \case{1}{2} s_{ij}s_{ij}$, corresponds to the
distortional energy. The material yields as the distortional energy
exceeds a material-dependent threshold $\sigma_{_{\rm Y}}^2$. The
fact that we treat this threshold as a constant, independent of the
state of deformation and its history, implies that we specialize for
``perfect'' plasticity. In 2-dimensions this yield condition reads
\cite{90Lub}
\begin{equation}
J_2
=\frac{\sigma^2_{1}-\sigma_{1}\sigma_{2}+\sigma^2_{2}}{3}=\sigma^2_{_{\rm
Y}} \ .
\label{Mises}
\end{equation}
Here $\sigma_{1\!,2}$ are the principal stresses given by
\begin{equation}
\sigma_{1\!,2} = \frac{\sigma_{yy}+\sigma_{xx}}{2} \pm
\sqrt{\frac{(\sigma_{yy}-\sigma_{xx})^2}{4}+\sigma_{xy}^2} \ .
\label{PrincipalStress}
\end{equation}

In the purely linear-elastic solution the crack-tip region is where
high stresses are concentrated (in fact diverging near a sharp tip).
Perfect plasticity implies on the one hand that the tip is blunted,
and on the other hand that inside the plastic zone the Huber-von
Mises criterion (\ref{Mises}) is satisfied. The outer boundary of
the plastic zone will be called below the ``yield curve", and in
polar coordinates around the crack tip will be denoted $R(\theta)$.

\begin{figure}[here]
\centering \epsfig{width=7 truecm,file=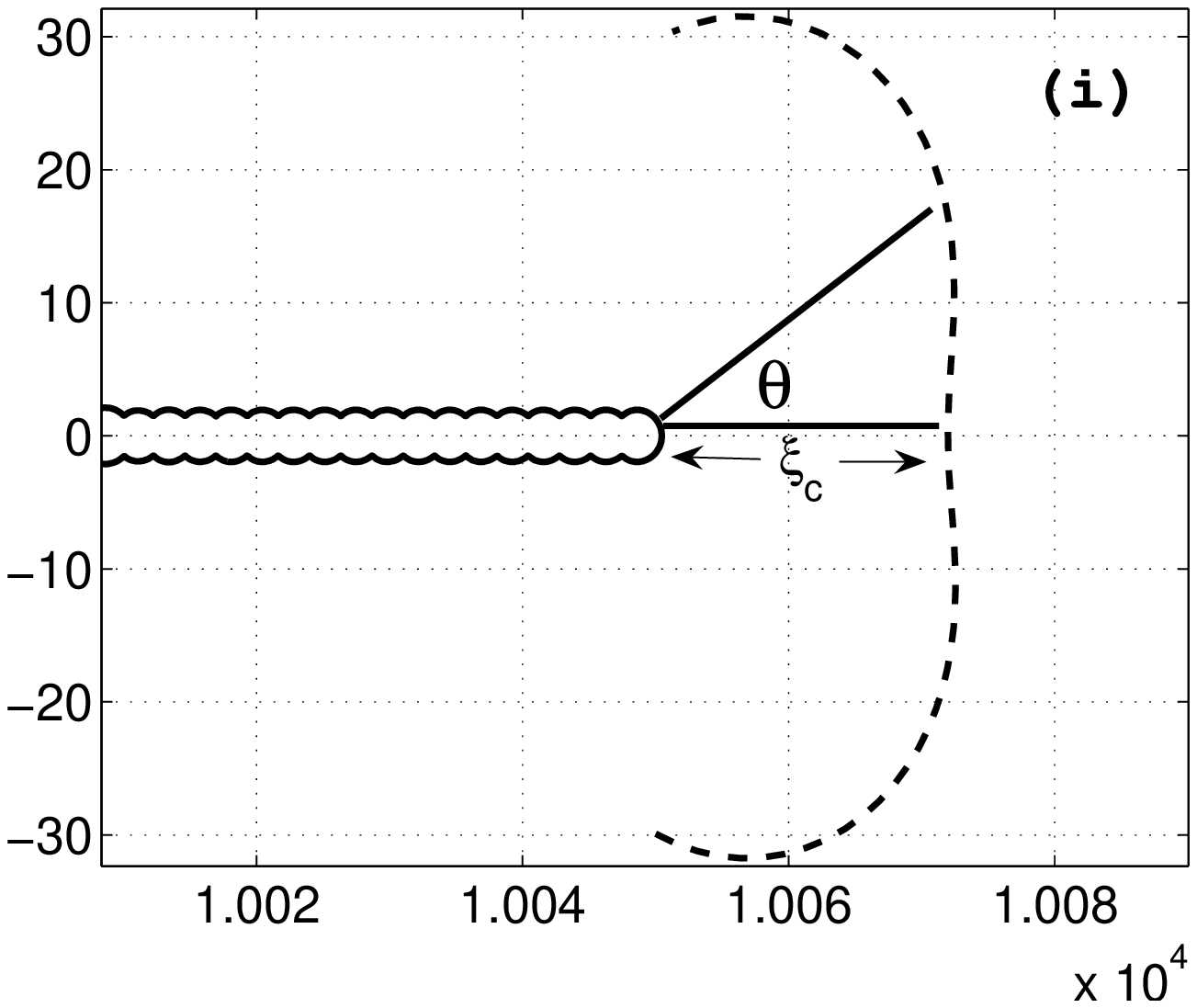}
\epsfig{width=7 truecm,file=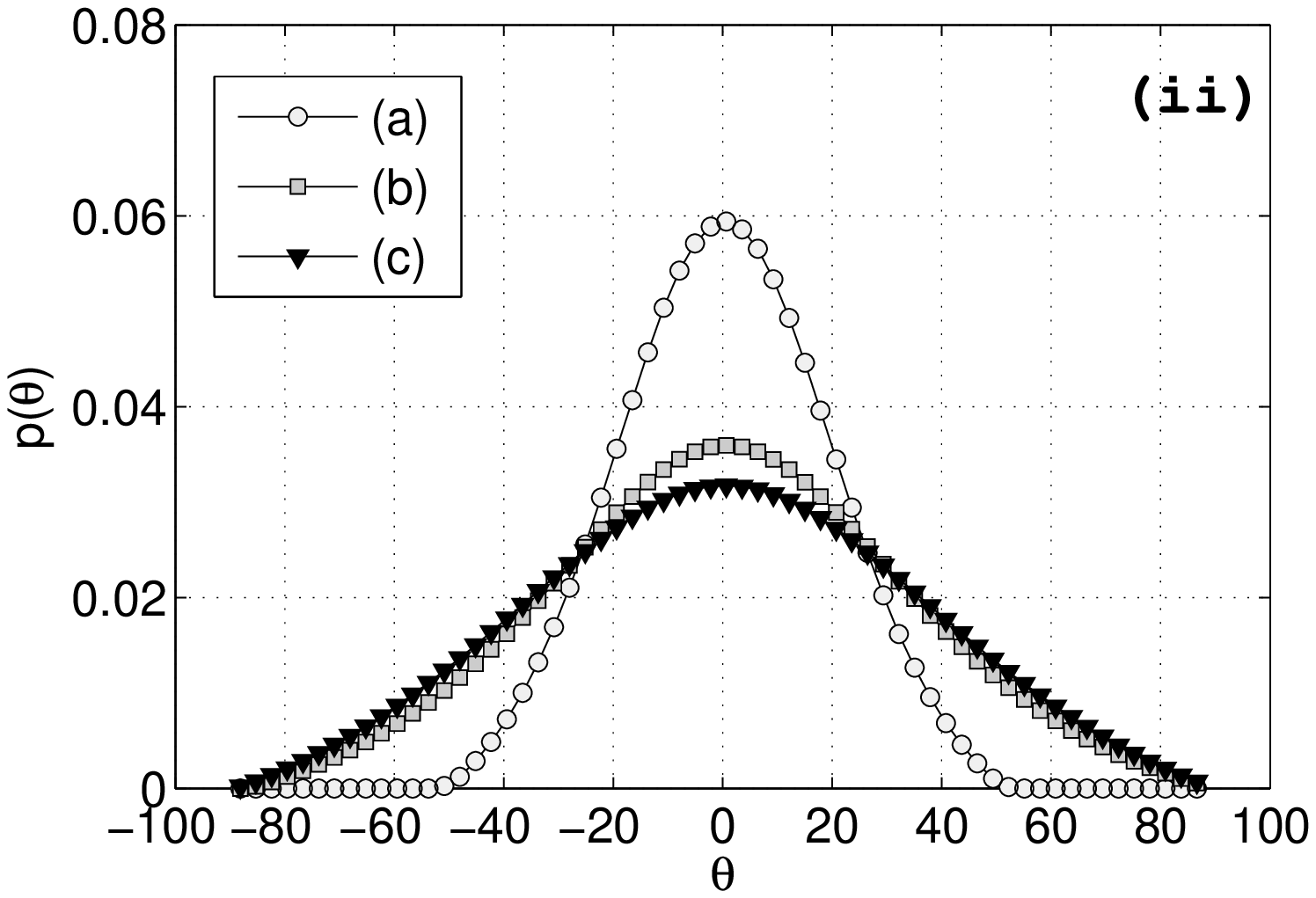 } \caption{Panel (i): the
tip of a straight crack and the yield curve in front of it.
\\ Panel (ii): three probability distribution functions calculated
for the configuration in (i). The abscissa is $\theta$, the angle
measured from the crack tip as seen in panel (i). The ordinate is
the normalized probability (per unit $\theta$) to grow in the
$\theta$ direction. The distributions are symmetric and wide enough
to allow deviations from the forward direction. For all the curves
$\frac{\sigma_{_{\rm Y}}}{\sigma^\infty} = 6$. For curve (a)
$\mathrm{p}(\theta) \propto \exp[(P-P_c)]-1$ and
$\frac{P_c}{\sigma^\infty}=8$, for curve (b) $\mathrm{p}(\theta)
\propto \exp[0.2 (P-P_c)]-1$ and $\frac{P_c}{\sigma^\infty}=6$ and
for curve (c) $\mathrm{p}(\theta) \propto P-P_c$ and
$\frac{P_c}{\sigma^\infty}=6$.} \label{straightsteps}
\end{figure}

Whatever is the actual shape of the
blunted tip its boundary cannot support normal components of the
stress. Together with Eq. (\ref{Mises})
this implies that on the crack interface
\begin{equation}
\sigma_{1} =  \sqrt{3}~\sigma_{_{\rm Y}}\!,~~~~~~\sigma_{2} = 0.
\label{PS}
\end{equation}
On the other hand, the linear-elastic solution, which is still
valid outside the plastic zone, imposes the outer boundary
conditions on the yield curve. Below we will compute the outer
stress field {\em exactly} for an arbitrarily shaped crack using
the recently developed method of iterated conformal mappings
\cite{03BMP}. For the present argument we will take the outer
stress field to conform with the universal linear-elastic stress
field for mode I symmetry,
\begin{equation}
\sigma_{ij}(r,\theta) =  \frac{K_{_{\rm I}}}{\sqrt{2\pi r}}
\Sigma^I_{ij}(\theta). \label{UF}
\end{equation}
For a crack of length $L$ with $\sigma^{\infty}$ being the tensile
load at infinity, the stress intensity factor $K_{_{\rm I}}$ is
expected to scale like $K_{_{\rm I}}\sim\sigma^\infty \sqrt{L}$.
Using this field we can find the yield curve $R(\theta)$. Typical
yield curves for straight and curved cracks are shown in the
insets of Figs. \ref{straightsteps} and \ref{pdf}.

\begin{figure}[top]
\centering \epsfig{width=.45\textwidth,file=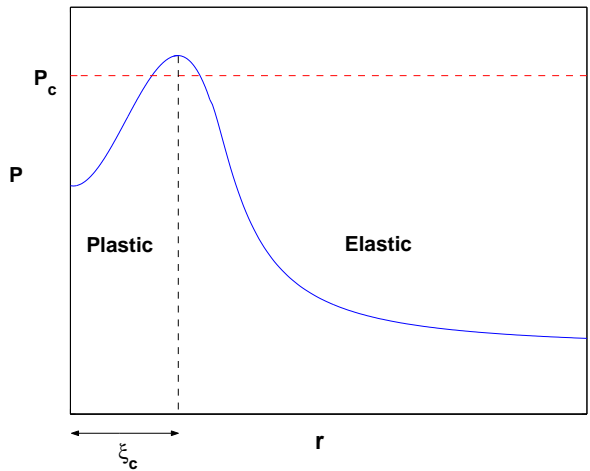}
\caption{A forward direction profile of the hydrostatic tension
$P$ in units of $\sigma_{_{\rm Y}}$. On the crack $P=\sqrt{3}/2$
and it attains a maximum of $\sqrt{3}$ on the yield curve. The
threshold line indicates a value of $P_c$ such that
$\case{\sqrt{3}}{2}\!\! <\! P_c \!\!<\!\! \sqrt{3}$. The typical
length $\xi_c$ is shown. Other directions exhibit qualitatively
similar profiles.} \label{Profile}
\end{figure}

The typical scale $\xi_c$ follows from the physics of the
nucleation process. It is physically plausible that void formation
is more susceptible to the growth of hydrostatic tension than to
distortional stresses. We assume that void nucleation occurs where
the hydrostatic tension $P$,  $P\equiv\case{1}{2}$Tr$\B \sigma$,
exceeds some threshold value $P_c$. The hydrostatic tension
increases when we go away from the tip and reaches a maximum near
the yield curve. To see this note that on the crack surface
$P=\case{\sqrt{3}}{2}\sigma_{_{\rm Y}}$ (cf. Eq. (\ref{PS})). On
the yield curve we use Eq. (\ref{UF}) and the Huber-von Mises
criterion together to solve the angular dependence of the
hydrostatic tension in units of $\sigma_{_{\rm Y}}$.  It attains a
maximal value of $\sqrt{3}\sigma_{_{\rm Y}}$ and is considerably
higher than $\case{\sqrt{3}}{2}\sigma_{_{\rm Y}}$ for a wide range
of angles. On the other hand the linear-elastic solution
(\ref{UF}) implies a monotonically decreasing $P$ outside the
yield curve. We thus expect $P$ to {\em attain its maximum value
near the yield curve}. This conclusion is fully supported by
finite elements method calculations, cf. \cite{85AM}. Finally,
since the nucleation occurs when $P$ exceeds a threshold $P_c$,
this threshold is between the limit values found above, i.e.
$\frac{\sqrt{3}}{2}\sigma_{_{\rm
Y}}\!\!\!<\!\!P_c\!\!<\!\!\sqrt{3}\sigma_{_{\rm Y}}$. The void
will thus appear at a typical distance $\xi_c$, see Fig. \ref{Profile}. An immediate
consequence of the above discussion is that $\xi_c$ is related to
the crack length via:
\begin{equation}
\xi_c \sim \frac{K^2_{_{\rm I}}}{\sigma^2_{_{\rm Y}}} \sim
\left(\frac{\sigma^{\infty}}{\sigma_{_{\rm Y}}}\right)^2 L \ .
\label{scaling}
\end{equation}

Note that $\xi_c$ is not a newly found length scale; it is the well
known scale of the plastic zone \cite{68Rice}. Its identification
with the cross-over length between two scaling behaviors of the
crack roughening is however new. This stems from the proposition
that positive correlations appear only between the positions of
nucleated voids. Below $\xi_c$ one enters the regime of plastic
processes whose theory is far from being settled. We should also
comment that it is possible that positive correlations appear even
below the scale of the plastic zone since experiments indicate that
several voids nucleate within the plastic zone \cite{99BP,03Cil}. We
should therefore consider the estimate in Eq. (\ref{scaling}) as an
upper bound on $\xi_c$. Finally, the fact that the plastic zone size
scales with $K_{_{\rm I}}$ as proposed in Eq. (\ref{scaling})
results from the assumption of perfect plasticity, i.e. that
$\sigma_{_{\rm Y}}$ is independent of the state of deformation and
its history. This is {\em not} true for real materials; usually
$\sigma_{_{\rm Y}}$ is not sharply defined; it can increase with
plastic deformations \cite{90Lub}. This phenomenon, known as
``work-hardening" or ``strain-hardening", might introduce other
dependencies on $K_{_{\rm I}}$ and include other length scales that
are related to the plastic deformations. We do not take such issues
into account in this simple model.

Naturally, the precise location of the nucleating void will
experience a high degree of stochasticity due to material
inhomogeneities. Since we do not know from first principles the
probability distribution for void formation, we consider in
our model below two possible distribution functions. In all
cases nucleation cannot occur if $P<P_c$. For
$P\!>\!\!P_c$ the void occurs with probability\\
\begin{eqnarray}
P&\propto& P-P_c\ , \label{Plin}\\
P&\propto& \exp[\alpha(P-P_c)]-1 \ . \label{Pexp}
\end{eqnarray}
In the exponential case we considered two different values of
$\alpha$. In Fig. \ref{straightsteps} we show three such pdf's as
they appear for a perfectly straight crack. We note that these
distributions are symmetric about the forward direction.
Nevertheless they have sufficient width to allow deviations from
forward growth. These deviations will be responsible later for the
roughening of the crack. For comparison examine also the pdf's for
a general crack which are shown in Fig. \ref{pdf}. There the
symmetry is lost: correlation to previous steps create a
preference for the upward direction. This source of positive
correlations is discussed below in greater detail.
\subsection{Crack Propagation}

Each growth step in our model is composed of two events.
Firstly the material yields near the crack
tip, creating a plastic zone with a void growing somewhere at the zone
boundary. Secondly the crack tip and the void coalesce. We
note that there is a separation of time scales between these two
events. The first is slow enough to be governed by a quasi-static stress field.
The second event occurs on a shorter time scale. It is clear that we forsake in the
one-void model any detailed description of the geometry on scales
smaller than $\xi_c$. Any relevant scaling exponent that will be
found in this model will refer to roughening on length scales larger
than $\xi_c$. In experiments it appears that several voids may
nucleate before the coalescence occurs, and in the
next section we will explore models with two voids ahead of the
crack. In the one-void model, the physical process in which the
crack coalesces with the multiple voids ahead of it is substituted
by a single void coalescence with the crack.

In spite of the simplification in dealing with only one void per
step, it was demonstrated in \cite{04BMP} that the one-void model
induces positive correlations between consecutive void
nucleations, leading eventually to an anomalous roughness exponent
larger than 0.5. Clearly, even this simple model requires strong
tools to compute the stress field around an arbitrarily shaped
crack, to determine at each stage of growth the location of the
yield curve and nucleating randomly the next void according to the
probability distributions discussed above. In a recent work we have
developed precisely the necessary tool in the form of the method
of iterated conformal mappings \cite{03BMP}.

In the method of iterated conformal mappings one starts with a
crack for which the conformal map from the exterior of the unit
circle to the exterior of the crack is known. (Below we start with
a long crack, in the form of a mathematical branch-cut of length
10000, and $\xi_c$ is of O(10)). We can then grow the crack by
little steps in desired directions, computing at all times the
conformal map from the exterior of the unit circle to the exterior
of the resulting crack. Having the conformal map makes the {\em
exact} calculation of the stress field (for arbitrary loads at
infinity) straightforward in principle and highly affordable in
practice. The details of the method and its machine
implementations are described in full detail in \cite{03BMP}. In
the next section we present the theory in great detail for the
two-void model, and avoid the repetition here.  We should just
stress that the method naturally grows cracks with tips of finite
curvature, and each step adds on a small addition to the tip, also
of a finite size that is controlled in the algorithm.

\begin{figure}[here]
\centering \epsfig{width=.45\textwidth,file=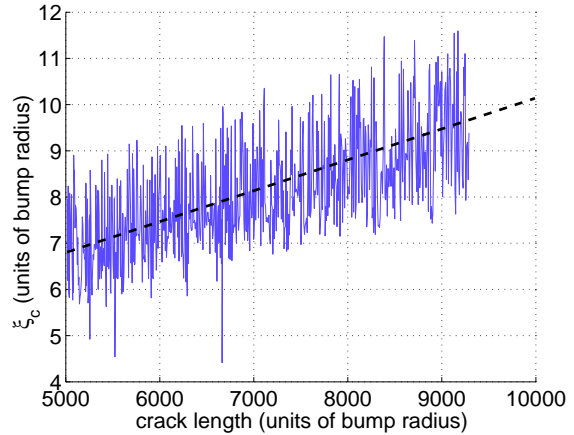}
\caption{The step size obtained using pdf (\ref{Plin}) (see section
\ref{onevoid}), vs. the crack length. Note that these results agree
with Eq.(\ref{scaling}) i.e. the step size grows linearly with the
crack length. The units here are the ``bump" radius which is
introduced explicitly in Eq. (\ref{phi}).} \label{stepsize}
\end{figure}

Having the stress field around the crack we can readily find the
yield curve and the physical region in its vicinity where a void
can be nucleated. Choosing with any one of the probability
distributions described above, we use this site as a pointer that
directs the crack tip. We then use the method of iterated
conformal mappings to make a growth step to coalesce the tip with
the void. Naturally the step sizes are of the order of $\xi_c$. In
Fig. \ref{stepsize} we present the actual step sizes as computed
with the pdf (\ref{Plin}), as a function of the crack length. The linearity
(in the mean) in $L$ is obvious. Note that the fluctuations about
the mean are strongly dependent on the pdf, and could in principle
be used to experimentally deduce the `correct' pdf by reverse
engineering. We reiterate that this model forsakes the details of
the void structure and all the length scales below $\xi_c$. Since
we are making {\em linear} steps below $\xi_c$, we anticipate
having an artificial scaling exponent $\zeta=1$ for scales smaller
than $\xi_c$. This is clearly acceptable as long as we are mainly
interested in the scaling properties on scales larger than
$\xi_c$.
\begin{figure}[top]
\centering \epsfig{width=.45\textwidth,file=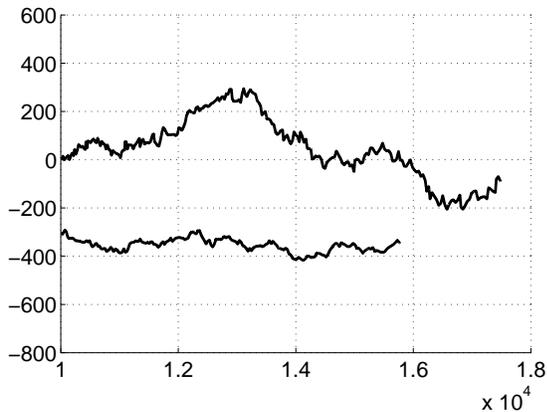}
\caption{Two typical cracks generated with our model. Note the
different scales of the abscissa and ordinate, and that the lower
crack had been translated by -300. The upper crack exhibits two decades
of self-affine scaling with a Hurst exponent $0.64$. The
lower crack has smaller standard deviation and therefore a shorter
scaling range. Nevertheless it appears that in its shorter scaling range
it exhibits an exponent that is very close to the upper crack.} \label{crack}
\end{figure}

\begin{figure}
\centering \epsfig{width=7 truecm,file=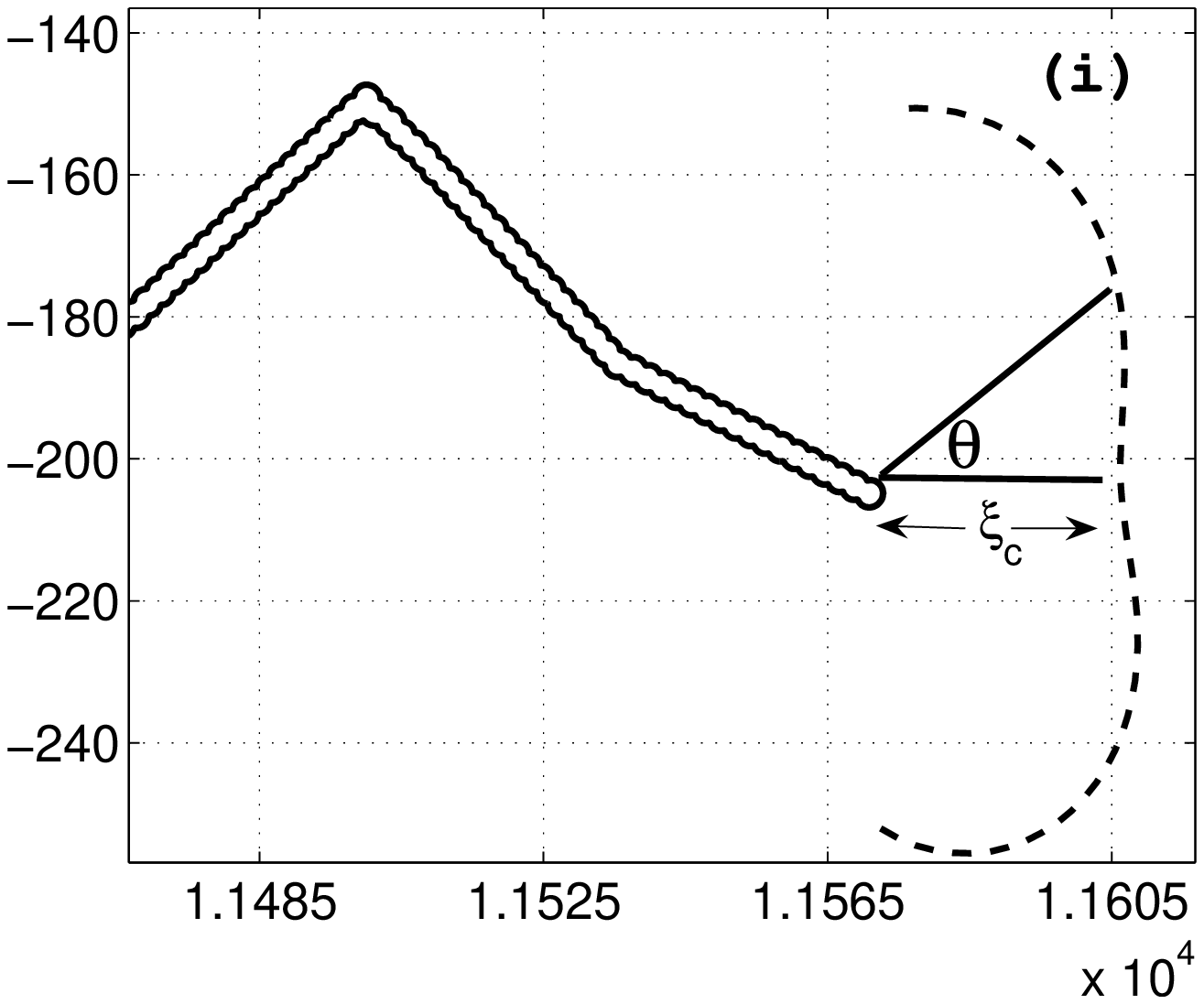 }
\epsfig{width=7 truecm,file=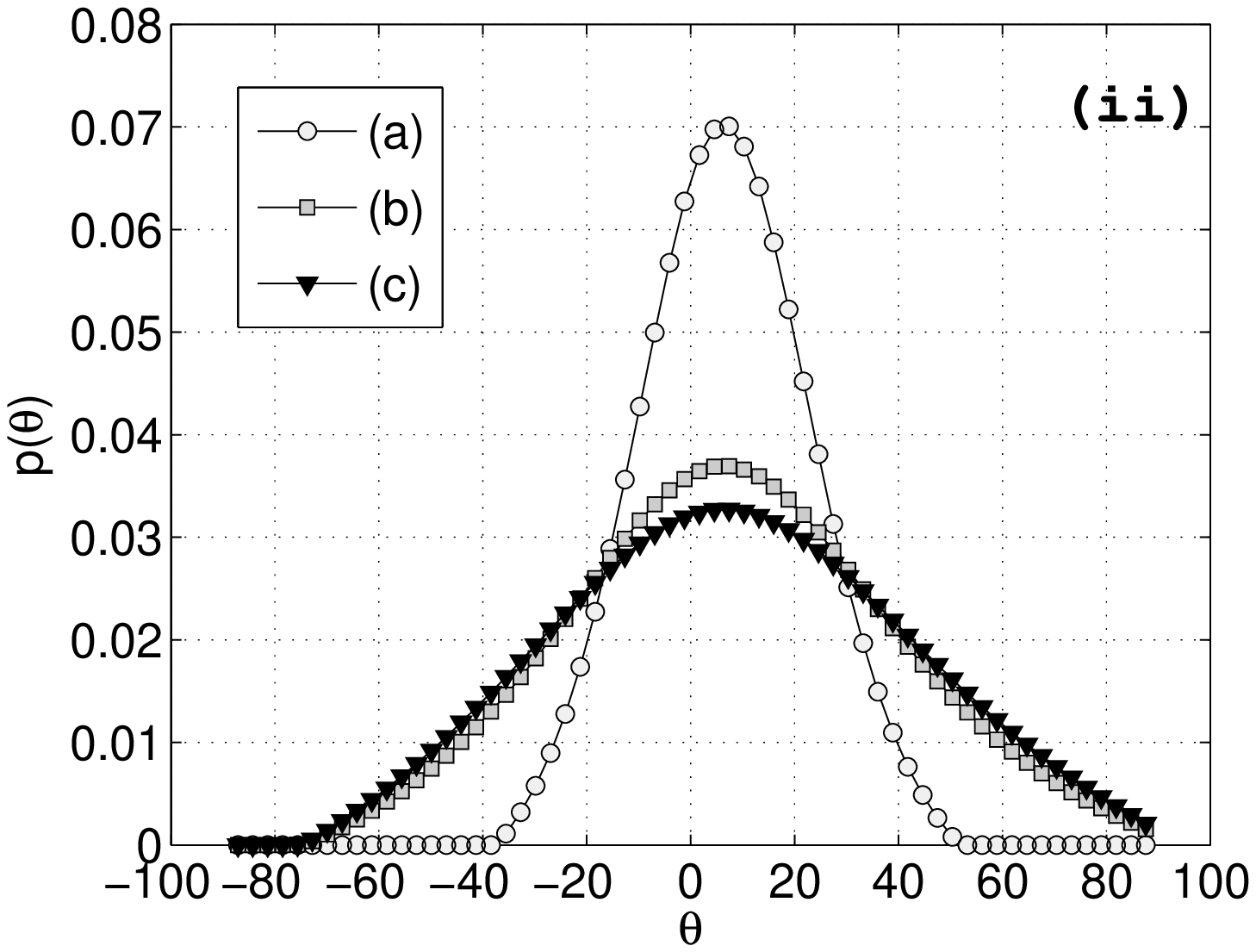 } \caption{Panel (i): the
tip of a ``rough" crack and the yield curve in front of it. Panel
(ii): three probability distribution functions calculated for the
configuration in panel (i). The abscissa is $\theta$, the angle
measured from the crack tip as seen in panel (i). The ordinate is
the normalized probability (per unit $\theta$) to grow in the
$\theta$ direction. The pdf's are those used in Fig.
\ref{straightsteps}, using the same parameters. Note the upward
preference in all the pdf's due to the broken symmetry.} \label{pdf}
\end{figure}

In Fig. \ref{crack} we present two typical cracks that were grown
using this method.  Both cracks were initiated from a straight crack
of length 10000. The upper crack was grown using the broader
exponential pdf of Fig. \ref{pdf} curve (b). The lower crack was
grown with the narrower pdf of Fig. \ref{pdf} curve (a). Clearly,
the upper crack exhibits stronger height fluctuations, as can be
expected from the wider pdf and the choice of parameters. For the
lower crack forward growth is much more preferred. In the upper
crack the positive correlations between successive void nucleation
and coalescence events can be seen even with the naked eye. This is
precisely the property that we were after. A neat way to see this
tendency is in the pdf's as they are computed on the yields crack
for a typical, rather than straight, crack. In Fig. \ref{pdf} we
show these pdf's for the crack whose yield curve is shown in the
upper panel. We see that now the symmetry of the pdf's is lost, and
positive values of $\theta$ are preferred. This is the source of
positive correlations that eventually give rise to the anomalous
roughening exponent.
 This is born out by
the measurements of the scaling exponent that we discuss next.

\begin{figure}[here]
\centering \epsfig{width=.45\textwidth,file=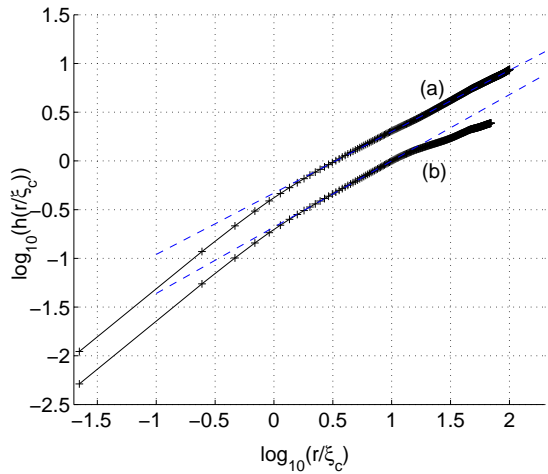}
\caption{Calculation of the anomalous roughening exponent. The
slopes of the dotted lines are $0.64$ for the upper plot (curve a) and 0.68 for the lower
(curve b). Note that the initial scaling
with slope $1$ is relevant for length scales smaller than $\xi_c$.
This scaling is unphysical, resulting from our algorithm that connects the crack
tip to a void by a straight line.} \label{h(r)}
\end{figure}

A quantitative measurement of the positive correlations is the
roughening exponent, that we compute as follows. Measuring the
height fluctuations $y(x)$ in the graph of the crack, one defines
$h(r)$ according to
\begin{equation}
h(r)\equiv\left<Max\left\{y(\tilde x)\right\}_{x<\tilde x<x+r}-
Min\left\{y(\tilde x)\right\}_{x<\tilde x<x+r}\right>_x \ .
\label{rough}
\end{equation}
For self-affine graphs the scaling exponent $\zeta$ is defined via the
scaling relation
\begin{equation}
h(r) \sim r^{\zeta} \ .
\end{equation}
In Fig. \ref{h(r)} we present a typical log-log plot of $h(r)$ vs.
$r$, in this case for the two cracks in Fig. \ref{crack} with
power-law fits of $\zeta=0.64$ and 0.68 respectively. Indeed as anticipated
from the visual observation of Fig. \ref{crack} the exponent is
higher than 0.5.
It turned out that all the cracks grown by our algorithm gave rise
to scaling plots in which a scaling range with $\zeta=0.66\pm 0.03$
is clearly seen. When the pdf allowed for a sizeable standard
deviation, the cracks gave a very nice scaling plot with at least
two decades of clear anomalous scaling. When the standard deviation
was small, the scaling range was more meager, as seen in Fig.
\ref{h(r)}.  It is interesting to stress that the anomalous scaling
exponent appears insensitive to the pdf used (although the extent of
the scaling range clearly depended on the pdf). We note that our
measured scaling exponents are very close to the exponents observed
in 2-dimensional experiments. (Of course we cannot expect a
2-dimensional theory to agree with 3-dimensional experiments - the
scaling exponents are, as always, dimension-dependent). In addition
the value of $\xi_c$ does not effect the scaling properties of a
crack, i.e. it doesn't seem to matter how long the step is, so long
as a wide distribution of angles is allowed.

Growing directly at the tip of the crack results in a very strong
preference for the forward direction, meaning that a step up will
most likely be followed by a step down, and vice versa, as shown
in \cite{02BLP}. The introduction of the physics of the plastic
zone results in creating a finite distance away from the tip to
realize the next growth step. Another crucial issue is the
existence of long range correlations. Since this aspect was not
made clear so far, we turn now to a discussion of the origin of
power law correlations.
\section{Long Range Positive Correlations in Fracture}
\label{long range}
The fact that the cracks generated by our model appear
self-affine with Hurst exponent $\zeta > 0.5$ implies that the
physical mechanism underling the crack growth is a long range
positive correlation process. We can gain intuition about the origin
of the long range correlations
by employing some known analytic results. Consider a long mode I
straight crack spanning the interval $[-L,0]$. Suppose now that
the crack shape is perturbed by a small out of plane fluctuation
of the form $\epsilon \psi(x)$, where $\epsilon>0$ is small. In
the presence of the perturbation the crack attains a small shear
component $K_{_{\rm II}} \neq 0$ at its tip. A first order
perturbation analysis in the amplitude $\epsilon$ reveals that
\cite{03BHP}

\begin{equation}
 K_{_{\rm II}} \sim -\epsilon \int_{-\infty}^{0}
\frac{\partial_{x}\left[\sigma^{(0)}_{xx}(x,0)\psi(x)\right]}{\sqrt{-x}}
dx \ , \label{perturb}
\end{equation}
where the superscript in $\sigma^{(0)}_{xx}(x,0)$ refers to the
solution in the absence of a perturbation. Note that due to the
fact that the crack is long we could set the lower limit of
integration to $-\infty$. Let us consider a positive perturbation
$\psi(x)$ that is symmetric around $-r$ and decays to zero at a
typical distance $\delta$ from $-r$. Since in our configuration
$\sigma^{(0)}_{xx}(x,0)= -\sigma^{\infty}$ for $x < 0$, it is
straight forward to show that Eq. (\ref{perturb}) yields
\begin{equation}
 K_{_{\rm II}} \sim -\frac{\sigma^{\infty} \epsilon  \delta}{r^{3/2}}
 \ , \label{scale}
\end{equation}
for $r \gg \delta$. Note that $K_{_{\rm II}}$ is {\em negative}. In order to
understand the effect of the perturbation on the probability
distribution function for the next void nucleation in our model,
we recall that this probability is determined by the hydrostatic
tension $P\equiv\case{1}{2}$Tr$\B \sigma$. In the case of a pure
tensile stress, $K_{_{\rm II}} = 0$, $P$ is symmetric around
$\theta=0$ and the probabilities of nucleating a void at positive
or negative angles are the same. In the presence of a small
negative shear component, $K_{_{\rm II}} < 0$, this picture
changes. The maximal hydrostatic tension is obtained at
\begin{equation}
 \theta_{max} \sim -\frac{K_{_{\rm I}}}{K_{_{\rm II}}}\ . \label{max1}
\end{equation}
Since to first order in $\epsilon$ the mode I stress intensity
factor $K_{_{\rm I}} \sim \sigma^{\infty} \sqrt{L}$ is unchanged
\cite{03BHP}, we obtain that the peak of the distribution is
shifted from zero to
\begin{equation}
 \theta_{max} \sim \frac{ \epsilon  \delta}{L^{1/2}r^{3/2}} > 0\ .
\label{max2}
\end{equation}
This relation shows that as a result of a positive perturbation in
the crack shape at a distance $r$ behind its tip the probability to
nucleate a void at a positive angle relative to the forward
direction is {\em higher} than the probability to nucleate a void
at a negative angle. Moreover, this positive correlation is long
ranged, decaying as $r^{-3/2}$.

The presence of long range correlations is reassuring, since they
are a must for the existence of a roughening exponent larger than
0.5. Note however that the above result does not determine in any
direct way the numerical value of the roughening exponent itself.
The actual exponent results from the cumulative effect of many
height fluctuations, and at present we do not have an analytic
theory predicting the numerical value of this exponent. Contrary
to self similar fractal growth patterns, where the fractal
dimension $D$ can be computed from the knowledge of the first
Laurent coefficient of the conformal map \cite{00DLP}, in self
affine graphs it is not obvious how to extract the roughening
exponent from the properties of the conformal map. At present we
are bound to the laborious process of actually growing the crack
and measuring the exponent. Needless to say this is theoretically
unsatisfactory, and new ideas on this issue should be very
welcome.
\section{two-void models: Mathematical formulation}
\label{twovoids} In this section we address the physical process
of nucleation of a second void in front of the crack tip. To improve upon the one-void model we need to
develop techniques to compute the exact stress field around a
crack with one void ahead (or a crack with two voids ahead, etc.).
Clearly, methods based on conformal maps from singly connected
regions (i.e. the unit circle) cannot suffice for this purpose.
Since conformal map techniques from doubly or multiply connected
regions are far less familiar, and since the computation of stress
field around doubly connected regions is interesting by itself, we
present the necessary techniques in some detail. Considerable
attention will be paid to the accuracy and the efficiency of the
calculation. In subsection \ref{applyconfmap} we review the basis
of the relevance of conformal maps to the solution of the
bi-Laplace equation. This goes back to Mushkelishvili's
series-expansion method \cite{53Mus}. We extend Mushkelishvili's
method to doubly connected regions, using conformal maps from the
annulus to the required doubly connected region. In Appendix
\ref{A} we elaborate the case of two circular holes. This case is
solvable analytically using bipolar coordinates \cite{bipolar},
and therefore provides a unique testing ground for the precision
of our method in a limiting case.

In this section we solve for the stress field in
an unbounded planar doubly connected region. Such calculations exist in the
literature; one general method is
known as the Schwarz alternating method. In this method one solves
the simply connected problems successively and superimposes them
in such a way that the boundary conditions are satisfied when the
procedure has converged (see for example \cite{Shwartz}). We will assess
our method by comparing with an analytical solution for two equally sized
circular holes using bipolar coordinates \cite{bipolar}. Our
method is quite genearal, allowing us to solve for the stress field in any
doubly connected region by explicitly solving the elastic
equations in a doubly connected geometry. The advantages of our
method are that it allows freedom in choosing
the shapes of the boundaries and it allows calculation of the
stress field near highly singular shapes such as long cracks. The
method can be implemented to very high precision as will be
shown below.
\subsection{Equilibrium equations for the stress field in a doubly
connected infinite medium.}
The theory of elastostatic fracture mechanics in brittle
continuous media is based on the equilibrium equations for an
isotropic elastic body \cite{86LL}
\begin{equation}
\frac{\partial \sigma_{ij}}{\partial x_j}= 0 . \label{equil}
\end{equation}
For in-plane modes of fractures, i.e. under plane-stress or
plane-strain conditions, one introduces the Airy stress potential
$U(x,y)$ such that
\begin{equation}
\sigma_{xx}=\frac{\partial^2 U}{\partial y^2} \ ; \sigma_{xy}=-
\frac{\partial^2 U}{\partial x\partial y} \ ;
\sigma_{yy}=\frac{\partial^2 U}{\partial x^2} \ . \label{sigU}
\end{equation}
Thus the set of Eq. (\ref{equil}), after simple manipulations,
translate to a Bi-Laplace equation for the Airy stress potential
$U(x,y)$ \cite{86LL}
\begin{equation}
\Delta \Delta U(x,y) =0 \ , \label{bilaplace}
\end{equation}
with the prescribed boundary conditions on the crack and on the
external boundaries of the material. At this point we choose to
focus on the case of uniform remote loadings and traction-free
crack boundaries. This choice, although not the most general, is
of great interest and will serve to elucidate our method. Other
solutions may be obtained by superposition. Thus, the boundary
conditions at infinity, for the two in-plane symmetry modes of
fracture, are presented as
\begin{eqnarray}
&&\!\!\!\!\!\!\!\!\sigma_{xx}(\infty)=0\ ;
\sigma_{yy}(\infty)=\sigma^\infty\ ;
 \sigma_{xy}(\infty)=0\quad \!\!\text{Mode I}\label{mode1}\\
&&\!\!\!\!\!\!\!\!\sigma_{xx}(\infty)=0\ ; \sigma_{yy}(\infty)=0\
;
 \sigma_{xy}(\infty)=\sigma^\infty\quad \text{Mode II}\ . \nonumber
\end{eqnarray}
In addition, the free boundary conditions on both boundaries (of crack and void) are
expressed as
\begin{equation}
\sigma_{xn}(s)=\sigma_{yn}(s)=0  \ , \label{bcm12}
\end{equation}
where $s$ is the arc-length parametrization of the boundaries and
the subscript $n$ denotes the out-ward normal direction.

The solution of the Bi-Laplace equation can be written in terms of
{\em two} analytic functions $\phi(z)$ and $\eta(z)$ as
\begin{equation}
U(x,y)= \Re [\bar z\varphi(z)+\eta(z)] \ . \label{Uphichi}
\end{equation}
In terms of these two analytic functions, using Eq. (\ref{sigU}),
the stress components are given by
\begin{eqnarray}
\sigma_{yy}(x,y)&=&\Re [2 \varphi'(z)+ \bar
z\varphi''(z)+\eta''(z)]\nonumber\\
\sigma_{xx}(x,y)&=&\Re [2 \varphi'(z)-\bar
z\varphi''(z)-\eta''(z)]\nonumber\\
\sigma_{xy}(x,y)&=&\Im [\bar z\varphi''(z)+\eta''(z)].
\label{components}
\end{eqnarray}
In order to compute the full stress field one should first
formulate the boundary conditions in terms of the analytic
functions $\varphi(z)$ and $\eta(z)$. The boundary conditions Eq.
(\ref{bcm12}) can be rewritten,  using Eq. (\ref{sigU}), as
\cite{53Mus}
\begin{equation}
\partial_t\left[\frac{\partial U}{\partial x}
+i\frac{\partial U}{\partial y}\right]=0\ . \label{bcU}
\end{equation}
Where  $\partial_t$ is the tangential derivative along the
boundaries. Condition (\ref{bcU}) must hold for each of the two
boundaries separately. Note that we do not have enough boundary
conditions to determine $U(x,y)$ uniquely. In fact we can allow in
Eq. (\ref{Uphichi}) arbitrary transformations of the form
\begin{eqnarray}
\varphi &\rightarrow& \varphi +iCz+\gamma\nonumber\\
\psi &\rightarrow& \psi +\tilde\gamma \ , \quad\psi\equiv \eta'
\label{gauge}
\end{eqnarray}
where $C$ is a real constant and $\gamma$ and $\tilde\gamma$ are
complex constants. This provides five degrees of freedom in the
definition of the Airy potential. It is important to stress that
whatever the choice of the five freedoms, the stress tensor is
unaffected; see \cite{53Mus} for an exhaustive discussion of this
point. We will explain below how to take advantage of these
freedoms to make the formulation simpler. When the domain is
doubly connected, the traction-free conditions (\ref{bcU}) can be
written using (\ref{Uphichi}) as
\begin{equation}
\varphi(z) + z\overline {\varphi' (z)} + \overline {\psi(z)} = D_1
\quad for \quad z \in C_1  \ \label{8}
 \end{equation}
and
\begin{equation}
\varphi \left( z \right) + z\overline {\varphi '\left( z \right)}
+ \overline {\psi \left( z \right)}  =  D_2\quad  for \quad z \in
C_2 \label{9}
\end{equation}
where $C_1$ and $C_2$ are the two boundary curves, and $D_1$,
$D_2$ are complex constants that are eventually uniquely
determined by the solution of the problem. To proceed we represent
$\varphi(z)$ and $\psi(z)$ in Laurent expansion form. Note that
$\varphi(z)$ and $\psi(z)$ have a poles inside both the boundaries
and therefore do not have a Laurent expansion around infinity
which is valid everywhere in the complex plane. But for $|z|> R$
where $R$ is a radius that encloses both the boundaries the
following expansion is valid since $\varphi(z)$ and $\psi(z)$ have
no poles in this region.
\begin{eqnarray}
\varphi(z) &=&\varphi_1 z + \varphi_0
+\varphi_{-1}/z+\varphi_{-2}/z^2+\cdots \ , \nonumber\\
\psi(z) &=&\psi_1 z + \psi_0 +\psi_{-1}/z+\psi_{-2}/z^2+\cdots \ .
\label{Laurentpp}
\end{eqnarray}
This form is in agreement with the boundary conditions at infinity
that disallow higher order terms in $z$. Using the boundary
conditions (\ref{mode1}), we find
\begin{eqnarray}
 \varphi_1&=&\frac{\sigma^{\infty}}{4}\ ;\quad
 \psi_1=\frac{\sigma^{\infty}}{2} \quad  \text{ Mode I}  \
,\nonumber\\
 \varphi_1&=&0 \ ; \quad\quad \psi_1=i\sigma^\infty
\quad \text{ Mode II} \ . \label{p1p1}
\end{eqnarray}
Where one of the freedoms in (\ref{gauge}) was used to choose
$\varphi_1$ to be real, using the real constant $C$ in
(\ref{gauge}). The four remaining freedoms will allow us later on
to fix $\varphi_0$ and $\psi_0$ in a convenient way.

\subsection{Application of conformal maps.} \label{applyconfmap}

\begin{figure}[here]
\centering \epsfig{width=.5\textwidth,file=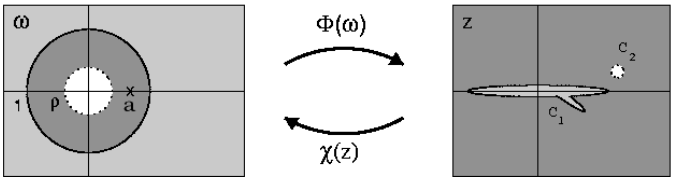}
\caption{Illustration of how the conformal map $\Phi(\omega)$
operates. The unit circle is mapped to $C_1$ and the inner circle
to $C_2$. The point $a$ is mapped to infinity.} \label{sketch}
\end{figure}

In order to enable the calculation of the stress field around an
arbitrarily shaped crack and a void, we conformally map the
annulus (having its outer radius set to one, i.e. the annulus is
$\rho<r<1$) onto the required doubly connected domain, see Fig.
\ref{sketch}. A well known fact is that simply connected domains
can be conformally mapped to any other simply connected domains
(relying on the Riemann mapping theorem). However, when dealing
with doubly connected domains there is an invariant quantity,
called the modulus (sometimes recasted as the extremal distance),
which is preserved under conformal mappings. As a result only
doubly-connected domains with the same modulus can be connected
via a conformal map. For an annulus the conformal modulus is just
the ratio of the inner radius and the outer radius, so that for
the $\rho<r<1$ annulus, the modulus is simply $\rho$. For that
reason the specific annulus which is taken as the domain to be
mapped onto the required crack+void domain cannot be just any
annulus, but has to be chosen correctly.

Suppose that we have such a conformal map $\Phi (\omega)$ (examples
are presented in the following subsections) that maps the
$\omega$-annulus domain onto the required physical $z$-plane with a
crack and a void in front of it. We employ this map to find the
solution of the stress field in the physical domain (we also
introduce the notation $\chi(z) \equiv \Phi^{-1}(z)$ for the inverse
map). Due to Eq. (\ref{Uphichi}), knowing the solution to the
Bi-Laplace equation in the annulus does not immediately provide the
solution as the Bi-Laplace equation in the physical domain through a
simple application of the conformal map, since in contrast to the
Laplace equation, the Bi-Laplace equation is not conformally
invariant. Nevertheless, the conformal mapping method can be
extended to non-Laplacian problems and provides a clear
simplification of the problem since the boundary conditions are much
easier to impose on a circular boundary than on the physical
boundary. We begin by writing our unknown functions $\varphi (z)$
and $\psi(z)$ in terms of the conformal map
\begin{equation}
\varphi \left( z \right) \equiv \tilde \varphi \left( {\chi \left(
z \right)} \right)\quad and\quad \psi \left( z \right) \equiv
\tilde \psi \left( {\chi \left( z \right)} \right)
\label{13}
\end{equation}
Now, $\tilde \varphi (\omega)$ and $\tilde \psi (\omega)$ are just
analytical functions in the annulus, apart from a simple pole
located at a point which is mapped to infinity in the $z$-plane,
cf. Fig. \ref{sketch}. Actually, a closer inspection leads us to
use the fact that in the $z$-plane $\varphi(z)$ and $\psi(z)$ have
truncated expansions as given in Eqs. (\ref{Laurentpp}). Thus, we
expect $\tilde \varphi(\omega)$ and $\tilde \psi (\omega)$ to be
of the general form
\begin{equation}
\tilde \varphi \left( \omega  \right) = A\Phi \left( \omega
\right) + \sum\limits_{n =  - \infty }^\infty  {\tilde \varphi _n
\omega
^n }
\label{14},
\end{equation}
and
\begin{equation}
\tilde \psi \left( \omega  \right) = B\Phi \left( \omega  \right)
+ \sum\limits_{n =  - \infty }^\infty  {\tilde \psi _n \omega ^n }
\label{15},
\end{equation}
where for Mode I fracture we can identify
\begin{equation}
A  = \varphi_1 = \frac{{\sigma _\infty  }}{4} \quad and \quad
B = \psi_1 = \frac{{\sigma _\infty  }}{2} \label{16},
\end{equation}
just like in Eq. (\ref{p1p1}) (from here on we use only Mode I
boundary conditions at infinity). In contrast to simply connected
domains, here we have positive as well as negative powers of
$\omega$ in the expansion. At this point we use the four remaining
freedoms ($\gamma$ and $\tilde{\gamma}$ in (\ref{gauge})) to
choose $\varphi_0$ and $\psi_0$ such that
\begin{equation}
\tilde{\varphi}_0=0 \quad and \quad \tilde{\psi}_0 = 0 \ .
\label{gauge1}
\end{equation}
To impose boundary conditions on the outer (unit) circle of the
annulus (i.e. $\left| \omega \right| = 1$) we write Eq. (\ref{8})
in the $\omega$ plane. This yields (using $\varepsilon =
e^{i\theta }$)
\begin{eqnarray}
 \frac{\sigma^\infty}{2}\big[\Phi(\varepsilon) &+&\overline {\Phi
( \varepsilon )}\big]  + \sum\limits_{n =  -
\infty }^\infty  {\tilde \varphi _n \varepsilon ^n }  +  \label{17}\\
  &+& \frac{{\Phi \left( \varepsilon  \right)}}{{\overline
{\Phi '\left( \varepsilon  \right)} }}\sum\limits_{n =  -
\infty}^\infty
{n\overline{\tilde  \varphi _n} \varepsilon ^{ - n + 1}
} + \sum\limits_{n =  - \infty }^\infty  {\overline{\tilde\psi _n}
\varepsilon ^{ - n} }  = D_1\nonumber.
\end{eqnarray}
Imposing boundary conditions on the inner circle of the annulus
(i.e. $\left| \omega \right| = \rho$)   we write Eq. (\ref{9}) in
the $\omega$ plane. This gives (using $\omega  = \rho \varepsilon
= \rho e^{i\theta }$)
\begin{eqnarray}
&& \frac{\sigma^\infty}{2}\big[\Phi(\rho\varepsilon) +\overline {\Phi
( \rho\varepsilon )}\big] + \sum\limits_{n
=  - \infty }^\infty  {\tilde\varphi _n \rho ^n \varepsilon ^n }  +
\label{18}\\
\!\!&&+ \frac{{\Phi \left( {\rho \varepsilon }
\right)}}{{\overline  {\Phi '\left( {\rho \varepsilon } \right)}
}}\sum\limits_{n =  - \infty }^\infty  {n\overline{ \tilde\varphi
_n} \rho ^{n - 1} \varepsilon ^{ - n + 1} }  + \sum\limits_{n =  -
\infty }^\infty {\overline{\tilde\psi _n }\rho ^n \varepsilon ^{ -
n} } = D_2 \nonumber.
\end{eqnarray}
To proceed we must Fourier Transform the functions $\Phi(\omega)$
and $\Phi(\omega)/\overline{\Phi^{'}(\omega)}$ on the
boundaries of the annulus. We use the following notation:
\begin{eqnarray}
\Phi(\epsilon) &=& \sum_{n=-\infty}^{\infty}c_n^{out} \epsilon^n \
, \nonumber \\
\frac{\Phi(\epsilon)}{\overline{\Phi^{'}(\epsilon)}} &=&
\sum_{n=-\infty}^{\infty}b_n^{out} \epsilon^n
 \ , \nonumber \\
 \Phi(\rho \epsilon) &=& \sum_{n=-\infty}^{\infty}c_n^{in} (\rho
\epsilon)^n \
, \nonumber \\ \frac{\Phi(\rho \epsilon)}{\overline{\Phi^{'}(\rho
\epsilon)}} &=& \sum_{n=-\infty}^{\infty}b_n^{in} (\rho
\epsilon)^n
 \ .   \label{FT}
\end{eqnarray}
Here indices with `out' refer to the outer unit circle, and
indices with `in' correspond to the inner circle of radius $\rho$.
Note that the Fourier coefficients on the inner boundary are not
the same as those on the outer boundary, this is because
$\Phi(\omega)$ has a pole inside the annulus. Also note that
contrary to the singly connected case, the expansion of
$\Phi(\omega)$ goes to infinity in both positive and negative
directions. Inserting Eq. (\ref{FT}) into Eqs. (\ref{17}) and
(\ref{18}), and gathering powers of $\epsilon$ we obtain the
following infinite set of equations for the coefficients
$\tilde\varphi_n$, $\tilde\psi_n$ and the unknown constants $D_1$
and $D_2$.
\begin{widetext}
\begin{eqnarray}
\tilde{\varphi}_n + \overline{\tilde{\psi}_{-n}} +
\sum_{k=-\infty}^{\infty} k b^{out}_{k+n-1}
\overline{\tilde{\varphi}_{k}}
 &=& -\frac{\sigma^{\infty}}{2} \left( c_{n}^{out} +
\overline{c_{-n}^{out}} \right)+ \delta_{n,0}D_1 \ , n=-\infty, \ldots
,\infty \ ,
 \label{mathbcLH3}  \\
\tilde{\varphi}_n \rho^{n} + \overline{\tilde{\psi}_{-n}}
\rho^{-n} + \sum_{k=-\infty}^{\infty} k b^{in}_{k+n-1} \rho
^{n+2k-2}\overline{\tilde{\varphi}_{k}}
 &=& -\frac{\sigma^{\infty}}{2} \left( c_{n}^{in} \rho^{n} +
\overline{c_{-n}^{in}} \rho^{-n} \right) + \delta_{n,0}D_2 \ ,
n=-\infty,\ldots ,\infty   \ . \nonumber \label{mathbcLH5}\
\end{eqnarray}
\end{widetext} This set of equations is well-posed and can be
solved in various ways. The simplest method is a truncation scheme
in which one neglects higher order terms in Eqs. (\ref{14}) and
(\ref{15}), (taking only a finite subset of coefficients
$\tilde{\varphi_n}$ and $\tilde{\psi _n}$, $n \ne 0$) and just
enough equations for solving those coefficients (as well as $D_1$
and $D_2$). When more and more coefficients $\tilde{\varphi_n}$
and $\tilde{\psi_n}$ are taken this scheme converges to the exact
solution. The efficiency and rate of convergence of this simple
scheme will be examined below.

The calculation of the Laurent expansion form of $\tilde \varphi
(\omega)$ and $\tilde \psi(\omega)$ provides the solution of the
problem in the $\omega$-plane. Still, one should express the
derivatives of $\varphi(z)$ and $\eta (z)$ in terms of $\tilde
\varphi (\omega)$ and $\tilde \psi (\omega)$ and the inverse map
$\chi (z)$ to obtain the solution in the physical $z$-plane. This
is straightforward:
\begin{eqnarray}
\varphi'(z)&=&\tilde\varphi'[\chi(z)] ~\chi'(z)\nonumber\\
\varphi''(z)&=&\tilde\varphi''[\chi(z)] ~[\chi'(z)]^2+
 \tilde\varphi'[\chi(z)] ~\chi''(z)\nonumber\\
\eta''(z)&=& \psi'(z)=\tilde\psi'[\chi(z)] ~\chi'(z).
\label{19}
\end{eqnarray}
Upon substituting these relations into Eq. (\ref{components}) one can
calculate the full stress field for an arbitrary doubly connected
infinite region.

In Appendix \ref{A} we present the application of
this formalism to the case of two circular holes where the conformal
map $\Phi (\omega)$
and its fourier coefficients are known analytically.
In this case one achieves extremely accurate solutions that can
be used as testing grounds for the truncation method that is
always available even when the fourier coefficients of the
conformal map are not given analytically.
\subsection{Conformal map for arbitrary crack with circular void
ahead} \label{confmap}

Our next task is to find the conformal map $\Phi(\omega)$ from the
interior of the annulus to the exterior of an arbitrarily shaped
crack and a void near its tip. We construct this conformal map by
composing three auxiliary maps. The properties of the stress maps
are as follows:
\subsubsection {Auxiliary Map 1} \label{aux1}
\begin{figure}
\centering \epsfig{width=.40\textwidth,file=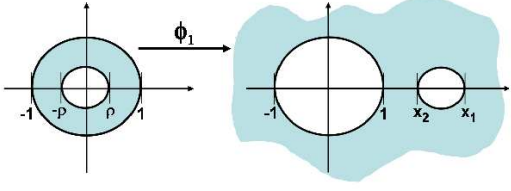}
\caption{Illustration of how the conformal map $\phi_1$ operates.
$\phi_1$ maps the annulus into the exterior of the unit circle and
a circle on its right. The unit circle boundary is mapped to
itself and the inner circle of radius $\rho$ is mapped to the
circle on the right with radius $\case{x_1-x_2}{2}$. }\label{phi1}
\end{figure}

Consider the map $\phi_1(\omega)$ given by,
\begin{eqnarray}
\phi_1(\omega) = \frac{a \omega - 1}{\omega-a}\ ,
\\ a\in\Re \quad\textrm{and} \ 0<a<1 \ .
\label{directphi1} \nonumber
\end{eqnarray}
$\phi_1(\omega)$ maps the annulus, i.e. $\rho<|\omega|<1$, onto
the exterior of the unit circle and an additional circle on the
right as is exemplified in Fig. \ref{phi1}. Note that the unit
circle is mapped onto itself and the inner circle is mapped onto
the circle on the right.
The map $\phi_1$ has two parameters, $a$ and $\rho$, the first appearing in its
definition (the point mapped to infinity), and the other in its domain of definition.
Both $a$ and $\rho$ are determined by the radius and the
center of the rightmost circle (see Fig. \ref{phi1}.) as follows:
\begin{eqnarray}
a &=&  \frac{x_1 + x_2}{1 + x_1 x_2 +
\sqrt{(x_{1}^{2}-1)(x_{2}^{2}-1)}} , \nonumber \label{a} \\
\rho &=& \frac{-1 + x_1
x_2-\sqrt{(x_{1}^{2}-1)(x_{2}^{2}-1)}}{x_1- x_2} \ , \label{rho}
\end{eqnarray}
where $x_1$ and $x_2$ are defined in Fig. \ref{phi1}. Note that
the following inequality must hold:
\begin{equation}
0 < \rho < a < 1  \ . \label{constraints}
\end{equation}
The inverse mapping is given by
\begin{eqnarray}
\phi_{1}^{-1}(z) &=& \frac{a z - 1}{z-a} \ .
\end{eqnarray}
Note that the inverse mapping is exactly the same symbol as the direct
one.
\subsubsection{Auxiliary Map 2}

The second map $\phi_2(\omega)$ is given by,
\begin{eqnarray}
\phi_2(\omega) &=& \omega \exp(i \theta)\ . \label{phi2}
\end{eqnarray}
$\phi_2(\omega)$ rotates the plane by an angle $\theta$
relative to the real axis. The inverse mapping is given by
\begin{eqnarray}
\phi_{2}^{-1}(z) &=& z \exp(-i \theta)\ .
\end{eqnarray}
\subsubsection{Auxiliary Map 3}

The role of $\phi_{3}$ is to map the exterior of the unit circle to
the
exterior of an arbitrary crack shape. Assume for now that we have such
a map at hand; in
subsection \ref{C} we present the explicit derivation of this map using
the tools
of iterated conformal map. At this point consider the composition of
all three maps.
\subsubsection{Composition of
the basic maps}
\begin{figure}
\centering \epsfig{width=.50\textwidth,file=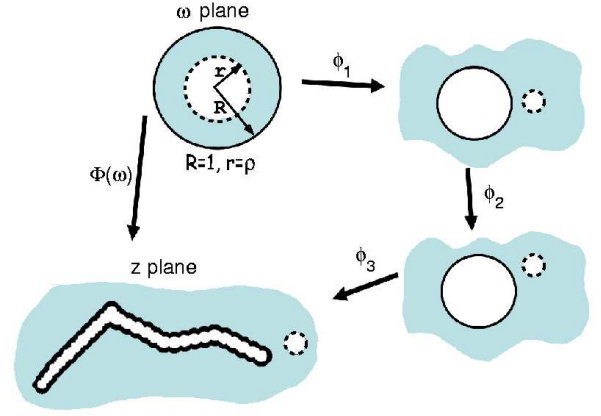}
\caption{Illustration of how the conformal map $\Phi(\omega)$
operates. $\phi_1$ maps the annulus onto the exterior of two
circles, $\phi_2$ rotates the whole plane and $\phi_3$ maps the
left circle into the desired crack shape, leaving the right circle
almost unchanged in its shape.}\label{compose}
\end{figure}

The desired mapping $\Phi(\omega)$ from the annulus
to the exterior of a crack and void is given by,
\begin{eqnarray}
\Phi(\omega) &=& \phi_3(\phi_2(\phi_1(\omega))) \ .
\label{composition}
\end{eqnarray}
The inverse map $\chi(z)$ is given by,
\begin{eqnarray}
\chi(z)&=&\Phi^{-1}(z) =
\phi_{1}^{-1}(\phi_{2}^{-1}(\phi_{3}^{-1}(z))) \ . \label{inverse}
\end{eqnarray}
The composition of the three auxiliary maps is illustrated in Fig.
\ref{compose}. First $\phi_1$ is applied and maps the interior of
the annulus into the exterior of two circles; then $\phi_2$ is
applied to allow the circle on the right to be rotated with
respect to origin. Finally, $\phi_3$ is applied to map the
exterior of the unit circle to the exterior of an arbitrary crack
shape. In total, $\Phi(\omega)$ maps the interior of the annulus to the
exterior of a crack and void, such that the outer boundary of the
annulus is mapped to the boundary of the crack and the inner
boundary is mapped to the void boundary. Notice that since
$\phi_{3}$ acts on the whole plane, it affects the void ,i.e. the
circle  on the right in Fig. \ref{compose}, as well as the crack
shape. An important property of the mapping we suggest is that for
all the configurations we are interested in, applying $\phi_{3}$
does not change the shape of the void in an appreciable way, i.e.
the void remains almost circular.

In order to create the mapping for a given crack and void
configuration one needs a set of points describing the crack's
path and the void's radius $R$ and center $z_0$. First
one constructs $\phi_3$ according to the desired crack shape (see
section \ref{C}). What is left is to obtain the values of $a$,
$\rho$ and $\theta$ (see auxiliary maps 1 \& 2). First we find
$x_1$, $x_2$ and $\theta$,
\begin{equation}
\theta = \arg(\phi_{3}^{-1}(z_0)) \ . \label{findparams1}
\end{equation}
\begin{eqnarray}
\frac{x_1 + x_2}{2} &=& |(\phi_{3}^{-1}(z_0))| \ , \nonumber \\
\frac{x_1 - x_2}{2} &=& |(\phi_{3}^{-1}(z_0 +
R)-\phi_{3}^{-1}(z_0) )| . \label{findparams2}
\end{eqnarray}
One can verify that using the above values for $x_1$, $x_2$
and $\theta$, a void with radius $R$ and center $z_0$ is obtained
in the $z$ plane. Substituting $x_1$ and $x_2$ in Eq. (\ref{rho})
we obtain the values of $\rho$ and $a$.

\subsubsection{Conformal Map from the unit circle to an arbitrary crack
shape}
\label{C}
\begin{figure}
\centering \epsfig{width=.45\textwidth,file=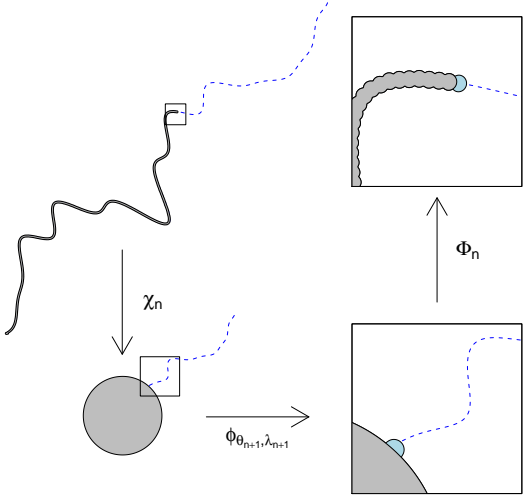}
\caption{Example of how to construct the conformal mapping along a
line.}\label{method}
\end{figure}

In this subsection we complete the identification of the auxiliary
map $\phi_3$. For the purpose of being reasonably self-contained we
reiterate here some aspects of the machinery of conformal maps. The
essential building block in the present application, as in all the
applications of the method of iterated conformal maps is the
fundamental map $\phi_{\lambda,\theta}$ that maps the exterior
circle onto the unit circle with a semi-circular bump of linear size
$\sqrt\lambda$ which is centered at the point $e^{i\theta}$. This
map reads \cite{98HL}:
\begin{eqnarray}
\label{phi}
   &&\phi_{0,\lambda}(w) = \sqrt w \left\{ \frac{(1+
   \lambda)}{2w}(1+w)\right. \\
   &&\left.\times \left [ 1+w+w \left( 1+\frac{1}{w^2} -\frac{2}{w}
\frac{1-\lambda} {1+ \lambda} \right) ^{1/2} \right] -1 \right \}
^{1/2} \nonumber\\
   &&\phi_{\theta,\lambda} (w) = e^{i \theta} \phi_{0,\lambda}(e^{-i
   \theta}
   w) \,.
   \label{eq-f}
\end{eqnarray}
The inverse mapping $\phi^{-1}_{\theta=0,\lambda}$ is of the form
\begin{equation}
  \label{eq:3}
  \phi^{-1}_{0,\lambda}= \frac{\lambda
z-\sqrt{1+\lambda}(z^2-1)}{1-(1+\lambda)z^2}z \ .
\end{equation}
By composing this map with itself $n$ times with a judicious
choice of series $\{\theta_k\}_{k=1}^n$ and
$\{\lambda_k\}_{n=1}^{n}$ we will construct $\Phi^{(n)}(\omega)$
that will map the exterior of the circle to the exterior of an
arbitrary simply connected shape. To understand how to choose the
two series $\{\theta_k\}_{k=1}^n$ and $\{\lambda_k\}_{k=1}^{n}$
consider Fig. \ref{method}, and define the inverse map
$\omega=\chi^{(n)}(z)$.  Assume now that we already have
$\Phi^{(n-1)}(\omega)$ and therefore also its analytic inverse
$\chi^{(n-1)}(z)$ after $n-1$ growth steps, and we want to perform
the next iteration. To construct $\Phi^{(n)}(\omega)$ we advance
our mapping in the direction of a point $\tilde z$ in the
$z$-plane by adding a bump in the direction of $\tilde
w=\chi^{(n-1)}(\tilde z)$ in the $w$-plane. The map
$\Phi^{(n)}(\omega)$ is obtained as follows:
\begin{equation}
\label{conformal} \Phi^{(n)}(\omega) =
\Phi^{(n-1)}(\phi_{\theta_n,\lambda_n}(\omega )) \ . \label{iter}
\end{equation}
The value of $\theta_n$ is determined by
\begin{equation}
  \label{eq:1}
  \theta_n=\arg [\chi^{(n-1)}(\tilde z)]
\end{equation}
The magnitude of the bump $\lambda_n$ is determined by requiring
fixed size bumps in the $z$-plane. This means that
\begin{equation}
  \label{lambdan}
  \lambda_n = \frac{\lambda_0}{|{\Phi^{(n-1)}}' (e^{i
\theta_{n}})|^2}.
\end{equation}
We note here that it is not necessary in principle to have fixed
size bumps in the physical domain. In fact, adaptive size bumps
could lead to improvements in the precision and performance of our
scheme. We consider here the fixed size scheme for the sake of
simplicity, and we will show that the accuracy obtained is
sufficient for our purposes. Iterating the scheme described above
we end up with a conformal map that is written in terms of an
iteration over the fundamental maps (\ref{phi}):
\begin{equation}
  \label{phin}
\Phi^{(n)}(w)=\phi_{\theta_{1},\lambda_{1}}\circ\ldots\circ\phi_{\theta_n,\lambda_n}(w)
\ .
\end{equation}
For the sake of newcomers to the art of iterated conformal maps we
stress that this iterative structure is abnormal, in the sense
that the order of iterates in inverted with respect to standard
dynamical systems. On the other hand the inverse mapping follows a
standard iterative scheme
\begin{equation}
  \label{chin}
\chi^{(n)}(z)=\phi^{-1}_{\theta_{n},\lambda_{n}}\circ\ldots\circ\phi^{-1}_{\theta_1,\lambda_1}(z)
\ .
\end{equation}

The algorithm is then described as follows; first we divide the
curve into segments separated by points $\{z_i\}$. The spatial
extent of each segment is taken to be approximately
$\sqrt{\lambda_0}$, in order to match the size of the bumps in the
$z$-plane. Without loss of generality we can take one of these
points to be at the center of coordinates and to be our starting
point. From the starting point we now advance along the shape by
mapping the next point $z_i$ on the curve according to the scheme
described above. The resulting map $\Phi^{(n)}(w)$ is employed as $\phi_3$
above.

\section{two-void models: Results and Discussion}
\label{twovoidsres}

\subsection{Computational aspects and Precision}
\label{comp} The method presented in the previous section has two
stages which involve numerical approximation. The first is a
numerical Fourier transform of the conformal map (see Eq. \ref{FT}) and
the
second is the truncation scheme (see discussion after Eq.
(\ref{mathbcLH3})). Aside from these two steps the method is
analytical. In appendix \ref{A} we test the truncation
scheme (with no numerical Fourier transform) for the case of two
circles and find
it to be extremely precise. In the next subsection we present a
comparison of our method to another theoretical calculation for
the straight crack and void geometry. This comparison serves as a
testing ground for the numerical Fourier transform and truncation
scheme
combined.
\subsection{Stress Intensity Factor for Straight Crack and void}
\begin{figure}
\centering \epsfig{width=.37\textwidth,file=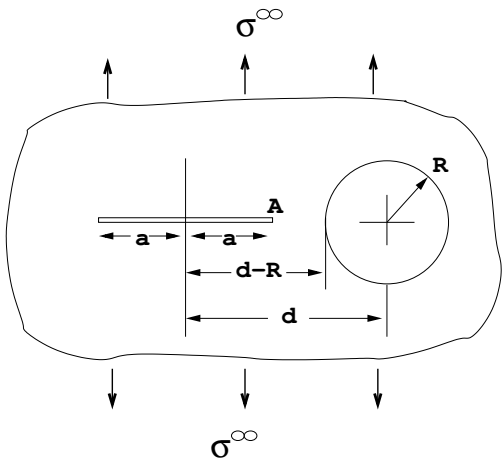}
\caption{The configuration of a line with a circle.  The
comparison of results exhibited in Fig. \ref {isidagraph} refers
to this configuration.}\label{isida}
\end{figure}
\begin{figure}
\centering \epsfig{width=.42\textwidth,file=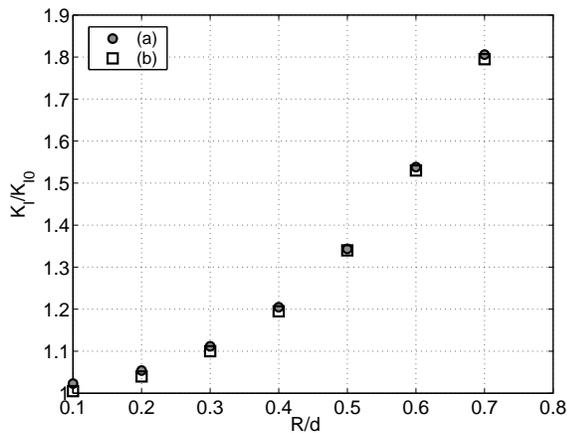}
\caption{$K_{_{\rm I}}$ for the configuration shown in Fig.
\ref{isida}. Note that the results are normalized by the stress
intensity factor with no void i.e $K_{_{\rm I0}} = \sigma^\infty
\sqrt{ \pi a}$. The stress intensity factor was calculated near the
point $A$ (i.e. at the tip which is close to the void). The
results were calculated for ${a}/(d-R)=0.4 $, varying
${d}/{R}$ in the range $[0.1,0.7]$. Isida's results (b) are
shown in squares, and ours (a) in circles. } \label{isidagraph}
\end{figure}

\label{stress intensity factor} The problem of a crack of length $2a$
in an infinite
domain, subjected to a remote uniaxial load
$\sigma_{yy}=\sigma^{\infty}$ and traction-free crack faces is
considered as the canonical problem in the theory of linear
elasticity fracture mechanics. The tensile stress component along
the tangent to the crack at the tip is given by \cite{Lawn}
\begin{eqnarray}
\sigma_{\varphi\varphi}(r,0)&=&\frac{K_{_{\rm I}}}{\sqrt{2\pi r}}
\ . \label{stress intensity factor1}
\end{eqnarray}
$K_{_{\rm I}}$ is known as the mode I stress intensity factor and
for a straight crack it is given by
\begin{eqnarray}
K_{_{\rm I}} &=& \sigma^{\infty} \sqrt{\pi a} \ . \label{stress
intensity factor2}
\end{eqnarray}
 Introducing a void in front of the crack causes an
increase in the stress intensity factor. The extent of this
increase depends on the void's distance from the crack and its
radius (see Fig. \ref{isida}). The problem of calculating the
stress intensity factor for a straight crack and void in front of
it has been solved using perturbational analysis by M. Isida
\cite{70Isi}. Using our method one can calculate the stress
intensity factor for the configuration in Fig. \ref{isida} by calculating
the stress in a region close to the crack tip and fitting it to
the form given in (\ref{stress intensity factor1}). A comparison
of our results and those of Isida is given in Fig.
\ref{isidagraph}. We note that Isida's results where extracted by
hand from a graph. Also, there are two small differences between
our geometry and that of Isida. First, the crack is not strictly a
branch cut but has a finite radius of curvature at the tip and
second, in our case the inclusion deviates slightly from a perfect
circular shape. All these factors together lead to an expected
difference of $\sim 2\%$ between the two methods. We conclude that
our results agree (to the expected precision) with those of
Isida's and are found to be accurate even in the vicinity of the
crack tip.

\subsection{Stress field near crack tip and void}
\label{stressres} 
\begin{figure}
\centering
\epsfig{width=7 truecm,file=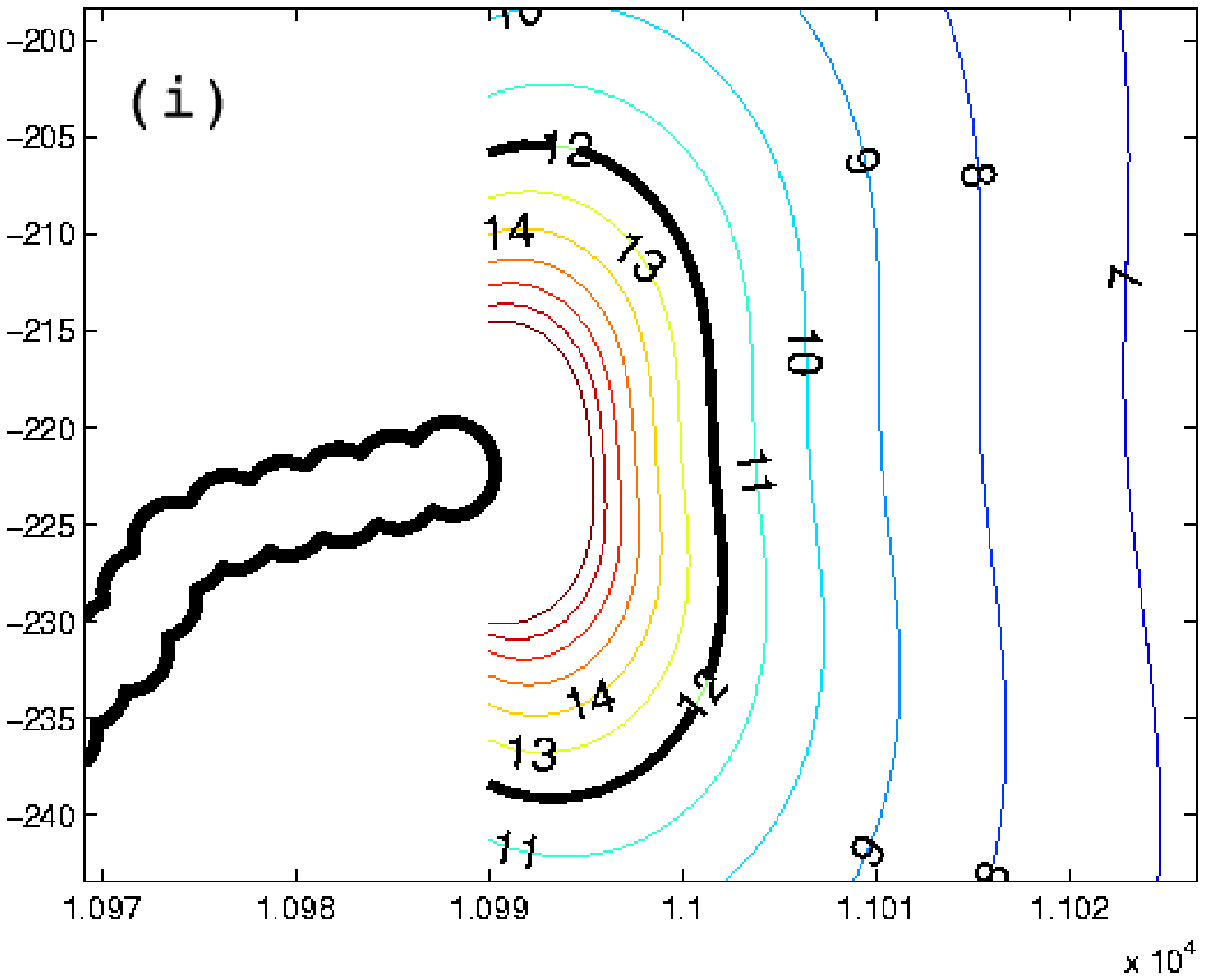}
\epsfig{width=7 truecm,file=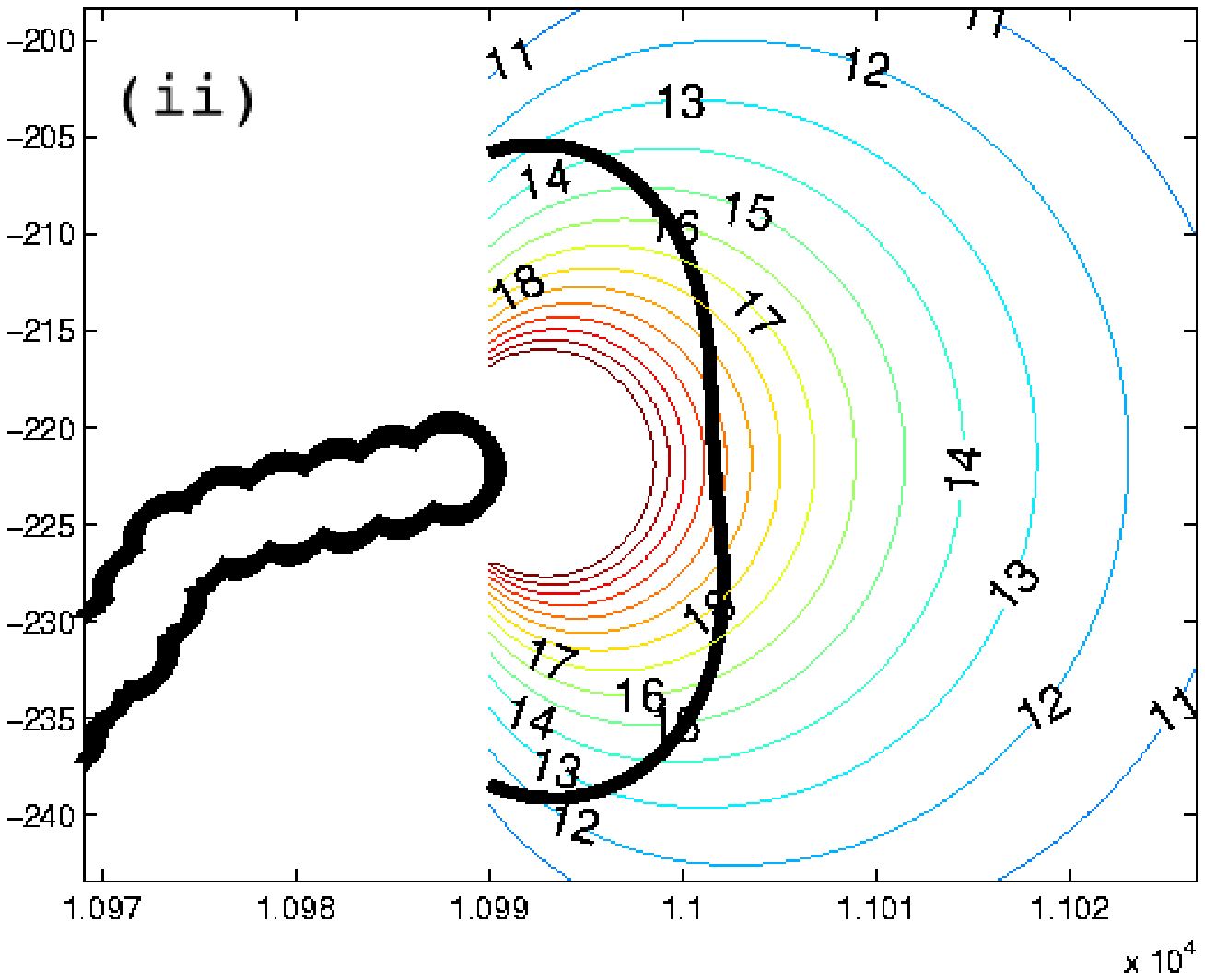}
\caption{ (color
online). Panel (i): field lines of $\sqrt{J_2}$, cf.
Eq.(\ref{Mises}), given in units of $\sigma^\infty$. The bold line
is the yield curve (i.e. $\sqrt{J_2} / \sigma^\infty = \sigma_{_{\rm
Y}} / \sigma^\infty = 12 $). The crack covers the $x$ interval
$[-10^4,1.01 \times 10^4]$ and the bump radius is $\sqrt{\lambda_0}
= 2$. The loading is Mode I with $\sigma_{yy}(\infty) =
\sigma^\infty$.\\  Panel (ii): field lines of the hydrostatic
pressure, $P \equiv \frac{1}{2} \mathrm{Tr} \B \sigma$ for the same
crack as the figure above, in units of $\sigma^\infty$. The bold
line is the \textsl{yield} curve from the previous figure. A
reasonable range for $P_c$ in such a configuration is $12 \le
\frac{P_c}{\sigma^\infty} \le 20$, in accordance with
$\frac{\sqrt{3}}{2}\sigma_{_{\rm
Y}}\!\!\!<\!\!P_c\!\!<\!\!\sqrt{3}\sigma_{_{\rm Y}}$ (see section
\ref{vonmises}). Note that $P_c$  in this range ensures forward
growth in the next step.} \label{novoid}
\end{figure}

\begin{figure}
\centering
\epsfig{width=7 truecm,file=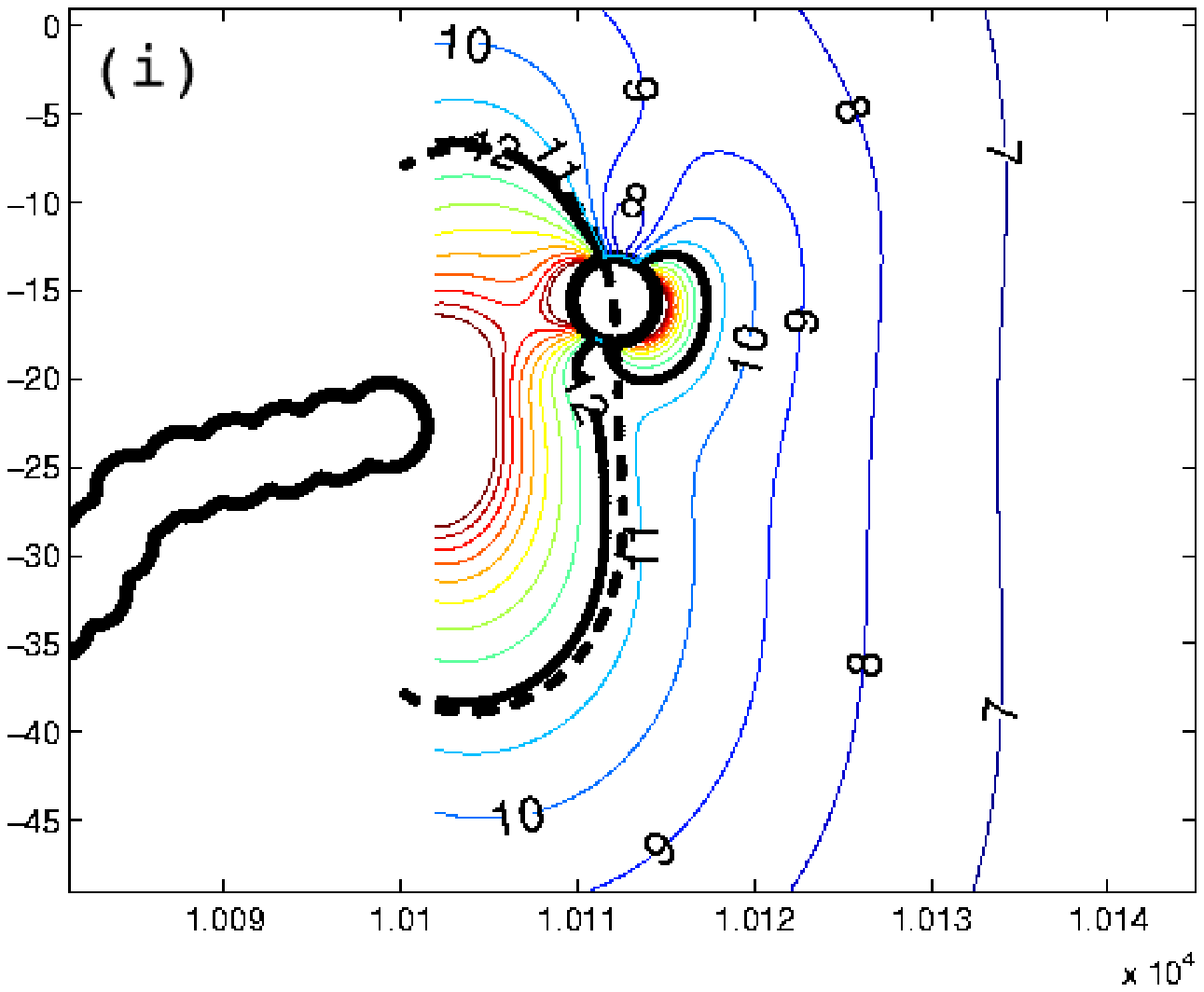}
\epsfig{width=7 truecm,file=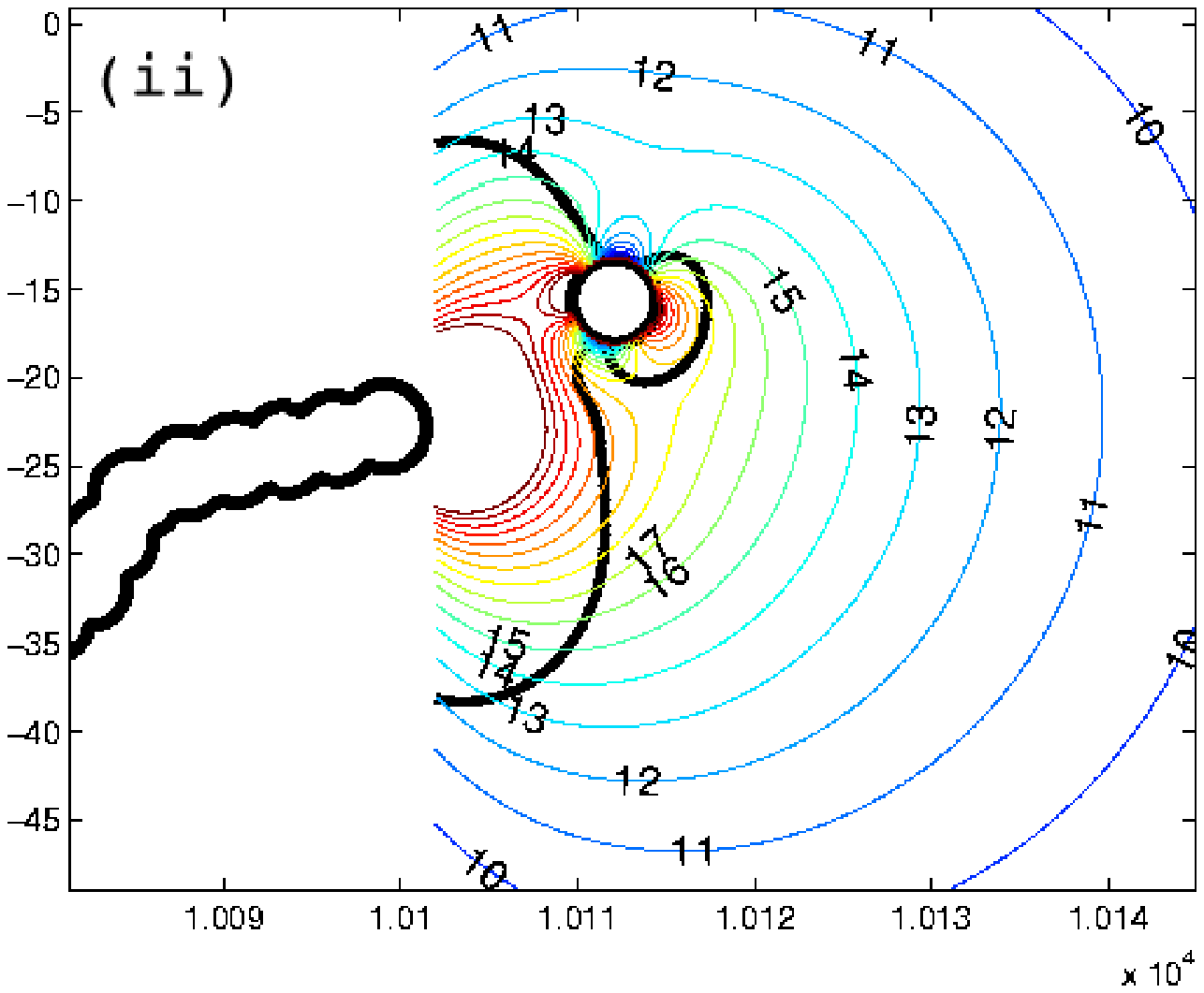}
\caption{ (color online)
The same crack as in Fig. \ref{novoid} with the addition of a void
on the yield curve.
\\ Panel (i): field lines of $\sqrt{J_2}$ in units of
$\sigma^\infty$. The bold line is the yield curve with the same
value of $\sigma_{_{\rm Y}}$ as in Fig. \ref{novoid} panel (i). The
dotted line is the yield curve from Fig. \ref{novoid} panel (i) i.e.
the yield curve for the same crack with no void. The addition of the
void
creates a very local perturbation of the yield curve.  \\
Panel (ii): field
lines of the hydrostatic pressure, $P \equiv \frac{1}{2} \mathrm{Tr} \B \sigma$ in
units of $\sigma^\infty$. The bold line is the yield
curve from the figure above. Again the perturbation of
the field lines with respect to Fig. \ref{novoid} panel (ii) is very
localized.} \label{withvoid}
\end{figure}

We now proceed to calculate the full stress
field for a configuration of a `rough' crack with a void. To
isolate the effect of the void on the stress field we first
calculate the stress for a crack without a void as before. We then
add a void and calculate the new stress field. The hydrostatic
pressure and yield stress without the void are shown in Fig.
\ref{novoid}. The same quantities after the void insertion are
shown in  Fig. \ref{withvoid}.

In theory we could now continue the void nucleation process by
inserting a new void in an appropriate point on the new yield
curve (the bold curves in Fig. \ref{withvoid}). It is obvious
however that this will gain us very little. The new yield curve is
only locally distorted by the presence of the void, and there is
definitely no typical distance of $2\xi_c$ that could be used to
create a decent two-void model.

The root of the problem is that we cannot determine the position
of the yield curve based on a fully elastic solution. The correct
position of the yield curve and the correct value of the stress on
it can only come from a solution of the full elastic-plastic
boundary value problem. When growing the first void we assumed
that the yield curve calculated from the solution of the fully
elastic problem is close to the elasto-plastic yield curve. This
assumption works well for the stress field of a crack alone, but
produces physically unacceptable results for the stress field of a
crack and void. Indeed, finite elements calculations that take
plastic flows into account \cite{85AM} indicate very clearly that
the new yield curve is further removed from the first void and
certainly does not coincide with the void boundary as it does in
our calculation.

The unavoidable conclusion is that the doubly connected conformal
calculation that is developed here is useful if one wants to
compute the stress field of an elastic material in which a hole
was inserted in the vicinity of a crack. It cannot be used however
to develop an approximate method of taking into account the
plastic yields that result in a successive appearance of two
voids. The first void can be inserted on the basis of elastic
calculations, but the second void cannot be added without a
considerably improved consideration of the plastic dynamics. As
long as the analytic aspects of plastic dynamics are not
elucidated better,  the one-void model is proposed as the best
available approach to roughening via growth with plastic
deformations.
\section{summary and conclusions}
The crack propagation model presented above, reproduces the
experimental appearance of self affine
crack rupture lines with an anomalous Hurst exponent. The long range correlations are created
by the stress field which satisfies boundary conditions on the
crack's  interfaces. The ability to solve the elastic boundary value
problem for an arbitrary crack is the basic building block of our
theory which enables us to capture these correlations.

To gain a sizeable scaling range with anomalous exponent our model
needs to employ a sufficiently wide probability distribution function in
the angle $\theta$ around the tip. While we could not discern a strong
dependence of the numerical value of the scaling exponent on the pdf, we
do observe a significant dependence of the {\em extent} of the scaling region.
Whenever we found a scaling range the numerical value of the exponent
was in the range of $\zeta=0.66\pm 0.03$.

We have also presented a general method for the calculation of the elastic
stress field surrounding a doubly connected region, like a crack and void at its tip, using
conformal maps. This method, although not useful for the creation
of a two-void model of crack growth, is quite
general and can be used for solving different physical problems
in doubly connected domains.

\acknowledgments

This work had been supported in part by European Commission under
a TMR grant, the Israel Science Foundation administered by the
Israel Academy, and by the Minerva Foundation, Munich Germany. Dr.
Katzav's research at the Weizmann Institute of Science is being
supported by the Edith and Edward F. Anixter postdoctoral
fellowship.

\appendix
\section{Two circular holes}
\label{A} In this section the specific case of two circular holes
is solved using the formalism of section \ref{twovoids}. For the
case of two equally sized circles there exists an analytical
solution \cite{bipolar}. In this solution bipolar coordinates are
used and a series expansion for the stress components is given in
which all the coefficients have a given closed form. We take
advantage of the bipolar method to check the precision of our
doubly connected method in the two circle limit.

\begin{figure}
\centering \epsfig{width=.50\textwidth,file=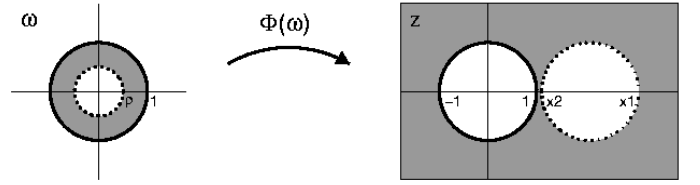}
\caption{An example of a conformal mapping of the annulus onto two
circular holes. In this example $x_1=1.1$ and $x_2=3.1$. Note that
the unit circle is mapped onto itself, while the $\rho$-circle in
the $\omega$-plane is mapped onto the circle on the right in the
z-plane.} \label{2circ}
\end{figure}

The conformal map for two circular holes is,
\begin{eqnarray}
\Phi(\omega) = \frac{a \omega - 1}{\omega-a}\ ,
\\ a\in\Re \quad\textrm{and} \ 0<a<1 .
\label{annulusmap} \nonumber
\end{eqnarray}
$\Phi(\omega)$ maps the annulus, i.e. $\rho<|\omega|<1$, onto the
exterior of the unit circle and an additional circle on the right,
see Fig. \ref{2circ}. Note that $\Phi(\omega)$ is the first
auxiliary map described in subsection \ref{confmap}. Since for
this case the conformal map has such a simple form, one need not
use a numerical FT to obtain its Fourier coefficients, therefore
the only approximation left in our method is the truncation of the
functions $\tilde{\varphi}(\omega)$ and $\tilde{\psi}(\omega)$ as
explained in subsection \ref{applyconfmap}. As a result the two
circle case allows us to isolate and estimate the error associated
with the truncation approximation by comparing with the exact
results of the bipolar method.

Specializing Eqs. (\ref{17})-(\ref{18}) for $\Phi(\omega)$ given
in Eq (\ref{annulusmap}) we obtain,
\begin{widetext}
\begin{equation}
 \frac{\sigma^\infty}{2}  \left( \frac{a \varepsilon  - 1}{\varepsilon
- a}
+ \frac{\varepsilon  - a}{a \varepsilon - 1} \right) +
\sum\limits_{n = - \infty }^\infty {\tilde{\varphi}_n
\varepsilon^n } + \frac{1}{1 - a^2 }\frac{\left( {a \varepsilon -
1} \right)^3 }{\varepsilon - a} \sum\limits_{n = - \infty }^\infty
{n {\tilde{\varphi}_n} \varepsilon ^{ - n - 1} } +
\sum\limits_{n =  - \infty }^\infty {  {\tilde{\psi}_n}
\varepsilon ^{ - n} }  = D_1 \ ,\label{22}
\end{equation}

and
\begin{equation}
 \frac{\sigma^\infty  }{2}  \left( \frac{a\rho \varepsilon  - 1}{\rho
\varepsilon  - a} + \frac{{a\rho  - \varepsilon }}{{\rho  - a\varepsilon
}} \right) + \sum\limits_{n =  - \infty }^\infty
{\tilde{\varphi}_n \rho ^n \varepsilon ^n } + \frac{{\left( {\rho
- a\varepsilon } \right)^2 }}{{1 - a^2 }}\frac{{a\rho \varepsilon
- 1}}{{\rho \varepsilon  - a}}\sum\limits_{n =  - \infty }^\infty
{n {\tilde{\varphi}_n} \rho ^{n - 1} \varepsilon ^{ - n -
1} } + \sum\limits_{n =  - \infty }^\infty
{{\tilde{\psi}_n} \rho ^n \varepsilon ^{ - n} }  =  D_2
\label{23}.
\end{equation}
\end{widetext}

As mentioned above, by expanding these equations in powers of
$\varepsilon$ one can get a well-posed set of equations which can
be solved by the truncation procedure. Using this method we
calculated the stress field surrounding two equally sized circles
see for example Fig. \ref{2circ}. We then solved the same problem
using the fully analytical bipolar method and found that both
methods give the same results to a very high degree of accuracy,
i.e. in all space the difference, $\Delta \sigma$, between the two
methods satisfies
\begin{equation}
\frac{\Delta \sigma}{\sigma^{\infty}} \simeq 10^{-14} \ .
\label{error}
\end{equation}
This holds in particular for area between the voids which might be
expected to give convergence problems.  We conclude that the
truncation method proves itself very accurate in this limiting
case.


\end{document}